\documentclass[twocolumn]{aastex6}

\fullcollaborationName{The Friends of AASTeX Collaboration}

\def\solm{M$_{\odot}\,$}
\def\msol{M$_{\odot}\,$}

\def\kms{km s$^{-1}$}

\def\solm{M$_{\odot}\,$}

\def\kms{km s$^{-1}$}

\def\casgm20{CAS-G-M$_{20}\,$}
\def\m20{M$_{20}\,$}

 \def\Mst{M$_{*}$}

 \def\Mtot{M$_{\rm{halo}}$}
 \def\Mvir{M$_{\rm{dyn}}$}
 \def\Mdyn{M$_{\rm{dyn}}$}

 \def\SK{S$_{0.5}$}
 \def\re{$R_{\rm{e}}$}
 \def\kms{\,km\,s$^{-1}$}
 \def\GALEX{\textit{GALEX}}
 \def\Spitzer{\textit{Spitzer}}
 \def\HST{\textit{Hubble Space Telescope}}
 \def\ACS{Advanced Camera for Surveys}
 \def\DEEP2{Deep Extragalactic Evolutionary Probe 2}

\begin{document}


\title{The Halo Masses of Galaxies to $z\sim 3$: A Hybrid Observational and Theoretical Approach}




\author{Christopher J. Conselice$^{1}$, Jonathan W. Twite$^{1}$, David P. Palamara$^{3,4}$, William Hartley$^{1,2}$}

\affil{Centre for Astronomy and Particle Theory, School of Physics \& Astronomy, University of Nottingham, Nottingham, NG7 2RD UK}

\altaffiltext{1}{University of Nottingham, School of Physics \& Astronomy, Nottingham, NG7 2RD UK}
\altaffiltext{2}{ETH Zurich, Institute for Astronomy, Wolfgang-Pauli-Strasse 27, CH-8093 Zurich, Switzerland}
\altaffiltext{3}{Monash University, School of Physics, Wellington Rd., Clayton, 3800, VIC, Australia}
\altaffiltext{4}{Monash Centre for Astrophysics (MoCA), Wellington Rd., Clayton, 3800, VIC, Australia}
\and


\begin{abstract}

We use a hybrid observational/theoretical approach to study the relation between galaxy kinematics and the derived stellar and halo masses of galaxies up to $z=3$ as a function of stellar mass, redshift and morphology. Our observational sample consists of a concatenation of 1125 galaxies with kinematic measurements at $0.4<z<3$ from long-slit and integral-field studies.  We investigate several ways to measure halo masses from observations based on results from semi-analytical models, showing that  galaxy halo masses can be retrieved with a scatter of $\sim 0.4$ dex by using only stellar masses.  
We discover a third parameter, relating to the time of the formation of the halo,  which reduces the scatter  in the relation between the stellar and halo masses, such that systems forming earlier have a higher stellar mass to halo mass ratio, which we also find observationally.  We find that this scatter correlates with morphology, such that early-type, or older stellar systems, have higher M$_{*}$/M$_{\rm halo}$ ratios.   We furthermore show using this approach, and through weak lensing and abundance matching, that the ratio of stellar to halo mass does not significantly evolve with redshift  at $1<z<3$.  This is  evidence for the regulated hierarchical assembly of galaxies such that the ratio of stellar to dark matter mass remains approximately constant since $z = 2$.   We use these results to show that the dark matter accretion rate evolves from $dM_{\rm halo}/d{\rm t} \sim 4000$ \msol year$^{-1}$ at $z \sim 2.5$, to a few 100 \msol year$^{-1}$ by $z \sim 0.5$. 
\end{abstract}






  \section{Introduction}

In the currently favored hierarchical picture for the formation of galaxies, small density 
fluctuations of  matter in the early universe induce the first dark matter halos to 
collapse (e.g., White \& Rees 1978; Davis et al. 1985).  Gas later collapses
 within these halos and eventually cools to form the first stars.  At the same time this occurs, these
dark matter halos are merging and accreting matter, and thereby grow 
in both dark matter and baryonic content over time.   The details of this picture have however yet to 
be worked out, and we are just
starting to understand the contribution of various processes responsible for galaxy assembly among many
others (e.g, Ownsworth et al. 2014; Mundy et al. 2017).    

Dark matter makes up a major portion of the total mass within the universe, yet due to 
observational constraints very little is known concerning how dark matter has evolved within 
galaxies over the history of the universe.  Measuring dark matter masses is however of fundamental 
importance to our full 
understanding of galaxy evolution/formation since dark matter halos drive their gravitational 
interactions as galaxies evolve through time (e.g., Conroy \& Wechsler 2009; Foucaud et al. 2010; 
Behroozi et al. 2013; Skibba et al. 2015).  Ideally we would like to be able to trace dark 
matter and its relationship to individual galaxies as a function of
cosmic time, yet this has proven difficult.

On the largest scales, 
massive dark matter halos hold galaxy groups and clusters together, helping to shape the 
largest  environments 
in the universe.  Predictions for the structure of these largest dark matter 'scaffoldings' 
match well with observations (Springel et al. 2005).  However, dark matter on the scales of galaxies 
is the ultimate way to test our cosmological ideas, as it is where
the very complex interplay between baryons, star formation, and active galactic nuclei all 
contribute to the structure and evolution
of galaxies.  Therefore tracing dark matter within galaxies over cosmic time is an important
observational goal, yet one that was until recently not possible to realize in any significant way.

The presence of dark matter within galaxies is ideally inferred from observations of baryonic 
matter if possible. The traditional method for measuring the dark matter content of galaxies is through 
their internal kinematics, and using this to derive 
their so-called dynamical masses.   This is in fact how dark 
matter in galaxies was first inferred (e.g., Faber \& Gallagher 1979), and remains the 
primary method for measuring the amount of dark matter on the scale of 
galaxies.    With the increase in the number of integral field units (IFUs) (e.g., SINFONI, KMOS, FLAMES/GIRAFFE) and higher resolution multi-object 
long-slit 
spectrographs (for example FORS, MOSFIRE, GMOS and DEEP2) on telescopes,  
kinematic measures are being obtained for increasing large numbers of high 
redshift galaxies (e.g., Pasquini et al. 2002, Eisenhauer et al. 2003; Bonnet et al. 2004; Epinat et al. 2009; F\"{o}rster Schreiber et al. 2009;
Buitrago et al. 2013;  Wisnoioski et al. 2015; Tiley et al. 2016; Price et al. 2016; Wuyts et al. 2016; Guerou et al. 2017; Ubler et al. 2017).  
Measuring the dark matter content with 
kinematics for large numbers of individual galaxies at redshifts 
beyond $z \sim 3$ is however still very difficult using current technological capabilities.  
Furthermore, the observed kinematics of these observations only probes the inner parts of 
galaxies and often does not reveal the total or halo masses of galaxies.

On the other hand, stellar masses are the most easily accessible type of mass in galaxies to measure, and
there has been a considerable amount of work measuring stellar masses for galaxies at 
both low and high redshift up to $z \sim 10$ (e.g., Bell et al. 
2003; Bundy et al. 2006; Mortlock et al. 2011, 2015; Duncan et al. 2014).     Stellar masses 
(\Mst) of galaxies are usually calculated from multi-wavelength observations (for example Bundy et al. 2006; Mortlock et al. 2011).  Stellar mass functions and stellar mass
co-moving volume densities have now been studied in detail, indicating a clear history of stellar 
mass growth, whereby around half of all stellar mass is in place by $z \sim 1$ (e.g., Drory et al. 2005; 
Conselice et al. 2007; Elsner et al. 2008, P\'{e}rez-Gonz\'{a}lez et al. 2008; Mortlock et al. 2011;
Muzzin et al. 2013; Ilbert et al. 2013; Duncan et al 2014; Mortlock et al. 2015). 

In this paper we provide a detailed investigation of how stellar and dark matter
masses can be measured up to $z \sim 3$ in a systematic way on individual systems. 
We investigate different 
methods of measuring the dynamical and halo masses of galaxies through kinematics and abundance 
matching.  While measuring the masses of distant galaxies is now done using methods such as 
clustering, weak lensing and abundance matching, these methods are all statistical
in nature and cannot easily be used to predict the halo masses of individual galaxies.
We describe whether and how these masses agree with each other at high redshift, and discuss evidence
for any evolution.  
We furthermore use these techniques and results to determine the accretion 
rate of dark matter into galaxies at $z < 1.2$ and compare with theoretical models.  By measuring the dark matter masses of individual galaxies we may be better able to 
directly use galaxy evolution to test fundamental features of the universe, such as dark matter and 
cosmological parameters, as well as better connect simulated dark matter halos with
observations (e.g., Conselice et al. 2014).  

As such, we present a new approach to tracing dark matter in galaxies
which relies on theoretical models.  This is not ideal as we would naturally want a purely observational
method.  However, using a model is necessary to obtain the dark halo masses for other methods as well,
including through clustering and abundance matching.  Furthermore, often simple equations relating the size and velocity are used to measure the dark matter of distant galaxies, yet these simple relations have not yet been tested.  We carry out an examination of these issues in this paper, and discuss the implications for  galaxy formation.

In the first part of this paper  we measure the different types of masses 
(stellar, dynamical and halo matter mass, hereafter \Mtot) for a sample of 432 
galaxies from the 
DEEP2 survey supplemented by other samples (\S 2).  We furthermore describe how to 
measure different
masses for these galaxies, namely the stellar mass (\Mst), 
the dynamical mass (\Mdyn), and the total halo mass (\Mtot) in \S 3.
Then, in \S 4 we use these to derive relationships between \Mst\, and \Mtot, 
and how these  evolve between redshifts.   In essence this
paper's conclusions are divided into theoretical results in \S 3, and observational 
results in \S 4.    In \S~5 
we discuss our results in their entirety, and the limitations inherent in our method, and 
what future work is needed to make progress in this direction, while \S 6 is a summary. 

Throughout the paper, we adopt the cosmology $\Omega_{m}=0.3$, $\Omega_{\Lambda}=0.7$ 
and $H_{0}=70$\,km\,s$^{-1}$\,Mpc$^{-1}$.  For all masses and star formation rates we 
use a Chabrier  initial mass function (IMF, Chabrier 2003), converting masses using other IMFs,
if necessary, in order to compare with other work.

  \section{Data \& Sample}

    \label{sec:Data}
    \subsection{Sample Overview}

      In this section we describe the several samples of galaxies which we use in this paper.   
The data we use  is all previously published.   
Our primary sample in which we establish the relationship between stellar and 
dark matter masses up to $z \sim 1.2$ are from the Palomar Observatory Wide-field IR (POWIR) 
overlap with the DEEP2 survey (e.g., Davis et al. 2003; Conselice et al. 2008a).  

We furthermore 
utilize stellar masses and derived dynamical masses from studies such as Conselice et al. (2005), 
Treu et al. (2005), Erb et al. (2006), Forster-Scheiber et al. (2009), Epinant et al. (2009), 
Buitrago et al. (2013), Beifiori et al. (2014) and Price et al. (2016) in what we call our 
secondary samples.    We
describe these different samples, their uses, and limitations below.  

\subsection{Primary Sample}
\label{sec:priSample}

We use as our primary sample galaxies within the POWIR/DEEP2 survey (e.g., Conselice et al. 2008a; 
Davis et al. 2003; Kassin et al. 2007) for which kinematic measurements, stellar masses, measured 
star formation rates, and resolved sizes from Hubble Space Telescope imaging are available.    The DEEP2
survey is a 2.3 deg$^{2}$ area spectroscopic survey with a galaxy selection done in magnitude and
colour.  The spectroscopy for DEEP2 was taken with the DEIMOS spectograph, and spanned the range 
6100-9100\AA\, and was carried out with the
1200 line mm$^{-1}$ granting, providing a high enough resolution such that internal kinematics can
be measured.    However, for this study we only use data in the Extended Groth Strip (EGS) field where
deep HST and stellar mass information is available as part of the AEGIS survey (e.g., Conselice et al. 2008a; Davis et al. 2007).

        \begin{figure*}
          \centering
	\vspace{-2cm}
          \includegraphics[angle=90, width=17cm]{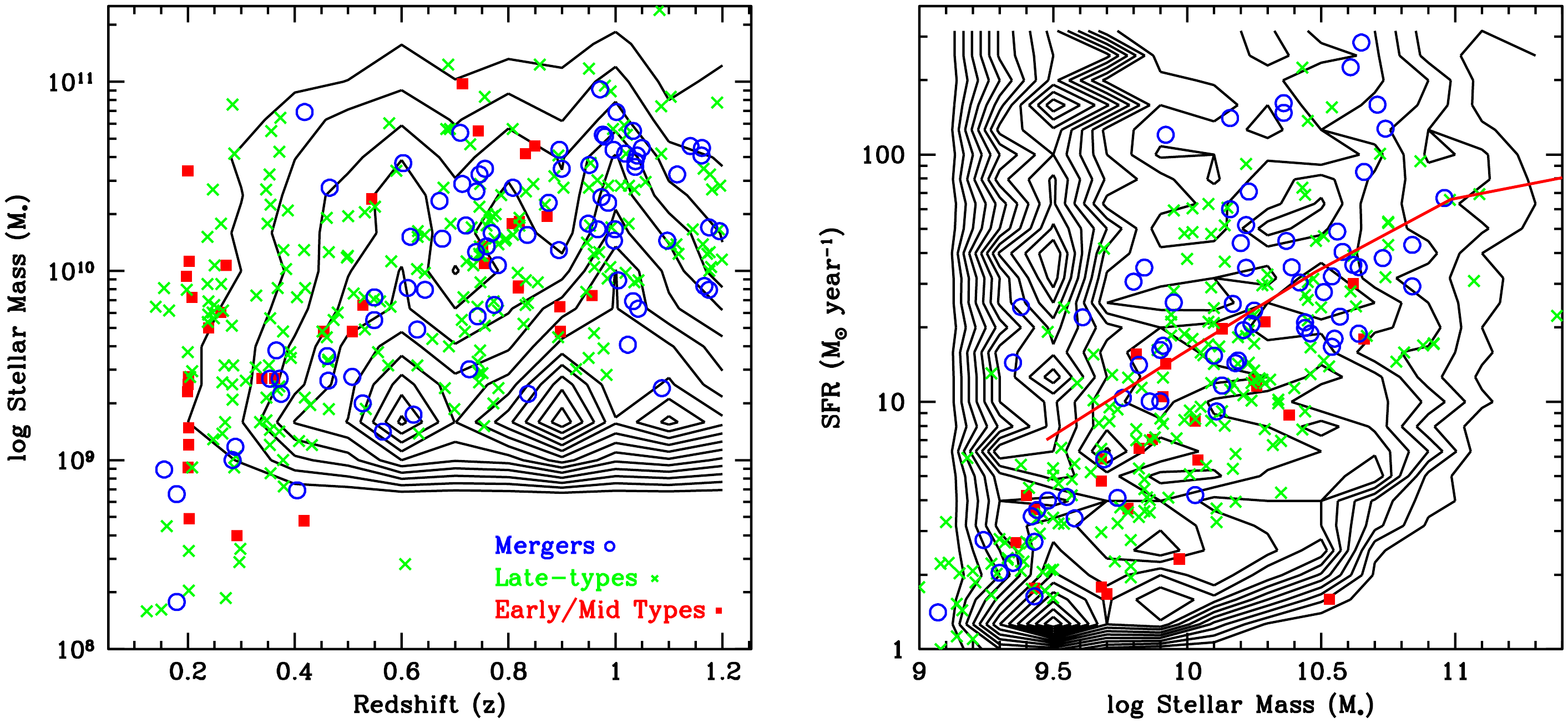}
	\vspace{-2cm}
          \caption{Plot on the left shows the stellar mass distribution as a function of redshift for our primary sample in reference to a field
galaxy sample down to a depth of K=26 AB.   The right panel shows the SFR
vs. stellar mass diagram.   Show as the colour points are the galaxies in our primary sample
of DEEP2/POWIR galaxies we use throughout this paper.   The contours are a sample of the Ultra Deep Survey sample galaxies taken from
the analysis of Mortlock et al. (2015).   The solid red line on the right panel is the main-sequence of star formation at $z = 1.5$ from Bauer et al. (2011).   The primary sample points are coded by the 
morphological type of each galaxy.  }

          \label{fig:BaryFrac}
        \end{figure*}

  The POWIR/DEEP2 primary sample spans a wide range in redshift, stellar mass, star formation, 
morphology and colour.  Using a redshift limit of $z = 1.2$ however biases the galaxies towards 
those with some star formation.  Our sample is selected based on an emission line cut, with 
integrated intensities 
$>1500\,{\rm e}^{-}\,$\AA$^{-1}$ in the summed one-dimensional spectrum.  The final sample after 
additional checks and cuts (see also Kassin et al. 2007), consists of 544 galaxies primarily 
selected on emission line strength and which are not AGN.  In order to use the measured kinematics
to derive further information, we discard galaxies which have velocity dispersions 
less than $\sim 10$\kms\ or rotational velocities less than $\sim 5$\kms.  This leaves 432 primary sample 
galaxies in total in which we investigate relationships between different types of masses.    We show the primary sample's stellar mass distribution with redshift
in Figure~1.

The DEEP2 primary galaxies fall within the POWIR survey (Conselice et al. 2007, 2008a), a large area ($\sim$1.5\,deg$^2$) deep NIR survey in the \textit{K} and \textit{J} bands incorporating multi-wavelength and spectroscopic data from other telescopes such as the CFHT, GALEX, Spitzer, and Chandra (Conselice et al. 2008a; Davis et al. 2007). We utilize the stellar mass catalogue from Conselice et al. (2007) to obtain the stellar masses of these galaxies from SED fitting as described in Bundy et al. (2006).

We furthermore also use single-orbit images from the \HST\ \ACS\ (ACS) in the F606W ($V$) and F814W ($I$) band-passes to classify these DEEP2 selected galaxies into morphological types through an automated morphological classification through the CAS method (Conselice 2003).  The details for how the CAS measurements are done are described in detail in Conselice (2003), Conselice et al. (2008) and Twite et al. (2011).   When we classify our galaxies we find that there is a mixture of types from ellipticals, disks, to irregulars/mergers within our sample.   The definitions for each morphological class are taken from Conselice, Rajgor \& Myers (2008b) and are summarized in Table~1. 

Quantitatively, our primary sample is composed of 7 (2 per cent) early-types, 36 (8 per cent) mid-types, 290 (67 per cent) late-types, and 99 (23 per cent) merging galaxies.  This is not representative of nearby galaxy types, but is closer to what one finds at higher redshifts where galaxies are almost all star forming 
(e.g., Mortlock et al. 2013).  However, the morphologies of these galaxies are not a significant part of our analysis, and we only use this aspect to test how morphology may affect the results we find.  Different morphologies are important when we examine the secondary sample later in the paper, which are selected by morphological type, (i.e.,
disk or elliptical) or through stellar mass.
   
The galaxies in this primary DEEP2/POWIR survey have star formation rates (hereafter SFRs) derived from UV measurements from \GALEX\ (Schiminovich et al. 2007, which also contains details on how the correction for dust attenuation was calculated), and/or IR measurements from \Spitzer\ 24$\mu$m data (Noeske et al. 2007).  Amongst this sample, 379 galaxies have a UV-derived SFR, and we use these values as an indicator of the star formation rate.  Another 23 only have IR-derived SFRs which are used; 30 galaxies have no SFR measurement.
        
It is important to note that our primary sample is biased in that we do not explore the entire 
galaxy population with a balanced mixture of morphologies, masses, and star formation rates.  
However, our sample is representative of all galaxy types  that exist at $z < 1$ (Figure~1),
with the exception of truly passive galaxies with no emission lines.   

We show this by examining where our primary sample falls in the 3-dimensional space of redshift, 
stellar 
mass and star formation rate for a stellar mass selected sample (Figure~1).   These properties are taken from the Ultra Deep
Survey (Mortlock et al. 2015) selected by $z < 1.2$, and using a K = 26 magnitude limit.  
As can be seen by the contours, we are primarily 
missing out on very low mass galaxies and passive systems without star formation.  Therefore 
any conclusions we draw in this paper based only on the primary sample are necessarily restricted 
to higher mass galaxies with some star formation.  We however supplement this sample with a broad range of galaxy types (\S 2.3).  Note, from Figure~1 that we do include galaxies 
with quite low star formation rates, although not at the highest masses.  However, we  
investigate these elliptical systems through our supplemental sample from Treu et al. (2005) 
and Beifiori et al. (2014) (\S 2.3), and other secondary sources.  

  The effective radii (\re) and Sersic index of the primary sample were measured by Conselice et al (2008a) and Trujillo et al. (2007) from the \textit{HST/ACS} F814W-band imaging of the POWIR EGS field.

        \begin{table}
          \begin{tabular}{l l l}
            \hline
            Morphological type & CAS parameters & Number \\
            \hline
            Early-type galaxies & $\mbox{C} > 21.5\log_{10}(A) + 31.2$ & 7 \\
            Mid-type galaxies   & $\mbox{C} < 21.5\log_{10}(\mbox{A}) + 31.2$ and & 36\\
                                & $\mbox{C} > 2.44\log_{10}(\mbox{A}) + 5.49$ and\\
                                & $\mbox{A} < 0.35$ \\
            Late-type galaxies  & $\mbox{C} < 2.44\log_{10}(\mbox{A}) + 5.49$ and & 290\\
                                & $\mbox{A} < 0.35$ \\
            Merging galaxies    & $\mbox{A} > 0.35$ & 99  \\
            \hline
          \end{tabular}
          \caption{CAS parameters for morphological classifications for our primary DEEP2/POWIR sample.  Also shown is the number of galaxies within each classification bin.  }
          \label{tab:CAS}
        \end{table}

      \subsection{Secondary Samples}
        \label{sec:secSample}

In our larger kinematic secondary sample, we use extensive data from various previous studies.  These
samples do not contain all the ingredients of the primary sample: kinematics, morphologies, sizes,
star formation rates, etc. so we only use them in a limited way to test how the results
derived from the primary sample apply.  The samples we use are:  Conselice et al. (2005), 
Treu et al. (2006), 
Erb et al. (2006), Epinat et al. (2009), F\"{o}rster Schreiber et al. (2009), Miller et al. (2011, 2014),
Buitrago et al. (2013), Beifiori et al. (2014), and Price et al. (2016). 

Part of this includes 257,000 galaxies from the 
BOSS survey which
are mostly massive early types (Beifiori et al. 2014). These are samples where there is at least 
kinematic and stellar mass measurements for galaxies at $z > 0.2$.  
This is not an exhaustive list of studies using internal 
kinematic data at high redshift, but it is representative of the different
types of galaxies that have been observed kinematically beyond the local universe.  
These secondary samples are often missing another component of information that we have in our primary sample.
This includes HST imaging for sizes, or 
accurate star formation rates, which limits our ability to obtain accurate  
dynamical masses as we can for the primary sample.

We give a brief summary of these samples here.   Conselice et al. (2005) measure the kinematics of 101 
galaxies at $0.2 < z < 1.2$ whose morphologies are disk-like.   A similar study by Miller et al. 
(2011, 2014) 
measures rotation curves for 129 disk-like galaxies at the same redshifts.   On the other hand, 
Treu et al. (2005) measure internal velocity dispersions of 165 elliptical and spheroidal galaxies at 
redshifts similar to those in these disk studies.  These are in contrast to the DEEP2 sample, which is 
not selected by any particular morphological type.
        
 For galaxies at the highest redshifts, we use results from studies by Epinat et al. (2009), and F\"{o}rster~Schreiber et al. (2009) who measure 
kinematics with integral field spectroscopy, and Erb et al. (2006) and 
Price et al. (2016) who use long-slit spectra.  The Erb et al. (2006) sample consists of 114 UV selected galaxies at $z \sim 2$ whose internal kinematics are measured with deep NIR long-slit spectroscopy.    The sample of F\"{o}rster~Schreiber et al. (2009) consists of 62 UV selected star forming galaxies at redshifts $1.3<z<2.6$ whose internal kinematics are measured with near-infrared IFUs.  Buitrago et al. (2013) measures kinematics for ten very massive galaxies with M$_{*} > 10^{11}$\,\msol at $z \sim 1.4$ using SINFONI on the VLT.   The Buitrago et al. (2013) galaxies are not selected by star formation rate or UV flux but by their high stellar masses.  The Epinat et al. (2009) sample consists of IFU kinematics for nine emission line galaxies also measured with SINFONI.  Finally, a more recent study of $z \sim 2$ galaxies with MOSFIRE on Keck contains a sample of 178 star forming galaxies (e.g., Kriek et al. 2015; Price et al. 2016).  

Occasionally within these samples the value of either the maximum velocity or the internal velocity dispersion is not available.  Usually this is the case when disk galaxies have no internal velocity dispersion measured, or ellipticals which have no rotational velocity. We infer the missing values for these by using the average value, at a given stellar mass, of the missing quantity for galaxies of similar morphology.  This is not needed for the majority of our sample, and our ultimate results do not depend on the exact replacement value used.

Details of these observations can be obtained through their respective papers.   It is important to point out that these samples were selected by a given property, usually morphology, or by having a high star formation rate. They are generally not representative of the galaxy population as a whole at their respective redshifts, nor are they homogeneous in terms of stellar masses.
        
      \subsection{Alternative Dark Mass Comparison Studies}
        \label{sec:compstudy}

One of the major goals of this paper is to investigate methods for measuring the dark matter halo masses of galaxies at $z>1$.  Measuring the halo masses of galaxies today is however not usually done through kinematics for high redshift galaxies. In fact, the most common way in which dark masses are measured are as a function of some property, usually stellar mass which is sometimes further divided into colour.  The halo masses of these samples are then measured through either clustering (e.g., Foucaud et al. 2010; McCraken et al. 2015; Skibba et al. 2015), lensing (e.g.,
Leauthaud et al. 2012), or more recently through abundance matching (e.g., Conroy et al. 2006; Behroozi 
et al. 2013).

  We compare our results with other studies that directly relate the stellar mass to the halo
mass.  This including using multi-epoch abundance matching (MEAM) from Moster et al. (2013).  We
also examine galaxy-galaxy weak lensing of galaxies from Leauthaud et al. (2012).   We
furthermore use the weak lensing results from van Uitert et al. (2016).  For clustering we
show the Foucaud et al. (2010) results comparing derived halo masses with stellar mass for 
stellar mass
selected samples.  These results are however averages for galaxies of a given type or stellar mass, and are not individual measures, as is potentially given by kinematic and size measurements.    We also compare our results of how halo and stellar masses relate with the same predictions from simulations (Springel et al. 2005; De Lucia \& Blaizot 2007; Benson
et al. 2012; \S 4).

 We discuss in more detail an alternative approach towards understanding the relationship between 
stellar and halo mass in \S 4.2 using abundance matching techniques which is becoming a 
popular method of tracing the stellar and halo mass evolution of galaxies.  

    \section{Observationally Derived masses}

In this section we explain how our various masses are measured for our primary and secondary samples, including how we measure their uncertainties.   Later in \S 4 we investigate  the halo masses of individual galaxies at high redshift without the use of strong gravitational lensing, or other direct ways to measure the halo masses of galaxies.    A list of these masses we use, and in what section of the paper they are defined and discussed are listed in Table~2.  In \S 4 we discuss masses which we derive using a hybrid of these observations and theory.  

      \subsection{Stellar masses}
        \label{sec:stellarMass}

The stellar masses of our primary DEEP2 sample galaxies are calculated in Bundy et al. (2006) and Conselice et al. 
(2007, 2008a) using targets with high quality spectroscopic or photometric redshifts from DEEP2 and 
using multi-wavelength photometry.  These masses are measured using a grid of 13,440 
synthetic SEDs from Bruzual \& Charlot (2003) spanning a range of exponential star formation histories, 
ages, metallicities and dust content.   These models are then fit to the photometry for each galaxy
to obtain a measure of stellar masses and other stellar population properties.    The stellar mass is 
determined by scaling the model \Mst/$L_{k}$ ratio to the measured K-band luminosity, $L_{k}$.  

The details of these masses is explained in Bundy et al. (2006) and Conselice et al. (2007, 2008), including the uncertainties.  While it is hard to pin down the total systematic errors on measurements of stellar masses, we have some idea based on the distribution of possible masses from different star formation histories. The typical total error is around 0.2 dex on these measurements.  We also examine how TP-AGB stars would change our measured masses, and found that the differences for our systems is very low, around 7\% on average (Conselice et al. 2007).   

We also calculate from radius measurements described in Trujillo et al. (2007), for the same galaxies in our primary sample, the stellar mass surface density as $\Sigma_{*}=\mbox{M}_{*}/\pi R_{\rm{e}}^{2}$.    We later use these surface densities to determine how density relates to kinematic properties and to the halo masses of our sample.

      \subsection{Dynamical Masses}
        \label{sec:dynMass}
     
In this section we discuss how we measure the dynamical masses of our galaxy 
samples.  This is an important question, and it is not necessarily a 
well-defined one.  Traditionally, a dynamical mass gives some measure of the 
dark matter mass within a galaxy. However, it is not obvious from an 
observational perspective how 
a galaxy's halo mass can be measured.  

We therefore divide our dark masses into two types - dynamical 
masses which are roughly the total mass of the galaxy within the observed 
portion of the galaxy which we discuss in this section, 
and the total halo mass, which must be derived based on inferring it from 
observables (\S 4.1).  In this 
section we only discuss observationally how to obtain a dynamical mass from 
observables - namely 
the size and internal velocities of galaxies.  

Note that there are many
ways to measure a 'dynamical' or 'kinematic' mass for galaxies.  What we
employ here is a method that is meant to be a measure of some fraction of
the total or halo mass, but is
not a total accounting for it.  In other words, what we use and define as a dynamical
mass is a kinematic indicator that scales with the halo mass, an idea we
test later in the paper.  This dynamical mass, by itself, does not reveal 
what the total or dark matter mass for galaxies is, and it has to be interpreted alongside the models by which its usefulness is derived.

 We measure this dynamical mass by first 
calculating 
a total `kinematic' indicator ($S_{K}$) for galaxies in our primary sample.
The value $S_{K} = K \times V_{\rm{rot}}^{2} + \sigma_{g}^{2}$ (Weiner et al. 2006; Kassin et al. 2007) is 
a quantity that combines measures of 
dynamical support from ordered motion with that from disordered motion by combining a factor for the 
maximal rotational velocity with the velocity dispersion of a galaxy. The value of K depends on the structure 
of the galaxy, whereby for systems that are spherically
symmetric and have an isotropic velocity dispersion with a density that declines as $\sim 1/r^{\alpha}$, 
then $\sigma = V_{\rm rot}/ \alpha^{1/2} = K^{1/2} \times V_{\rm rot}$.  The value K = 0.5 is a good 
compromise for a variety of values with the effective results unchanged if we used another slightly different value.
Assuming the quantities measured 
from the baryonic components of a galaxy trace the underlying total mass, this parameter is 
calculated for each galaxy in our primary and our secondary samples as:
        
        \begin{equation}
          S_{0.5}^{2} = 0.5 V_{\rm{rot}}^{2} + \sigma_{g}^{2}.
          \label{eqn:S05}
        \end{equation}
        
\noindent This quantity approximates the global internal kinematics of the galaxy-halo system (e.g., Binney \& Tremaine 1987).  This does however, imply certain assumptions must hold -- the system should be 
virialized, symmetric, and have an isotropic velocity dispersion, and an inverse power-law 
mass density distribution.  We use, as much as possible, the velocity dispersion  measurements at the same radius as the maximum velocity, both of which
are corrected for the effects of seeing (e.g., Conselice 2005; Kassin et al. 2007).

Although these assumptions could be considered approximately true for non-interacting galaxies, 
they would appear to be broken for disturbed and merging systems.  Kassin et al. (2007) however 
show that when using the \SK\ parameter to build stellar mass Tully-Fisher relations (\Mst TFR), 
the use of \SK\ instead of $V_{\rm max}$ has a significant effect for disturbed or compact 
galaxies and major-mergers, bringing them on to the same TF relation as other galaxies.  
 Combining these two kinematic properties together into the $S_{0.5}$ index 
can be interpreted as a measure of the underlying dynamical mass.  We discuss this in more detail in the appendix and in \S 5.1.1.    This implies that the \SK\, vs. stellar mass TFR is a more fundamental relationship linking the stellar mass to galaxy dynamics compared to the $V_{\rm{rot}}$ stellar mass TFR (\S 5.1.1). 

Assuming that the halo mass can be correctly measured from the velocity dispersion of a 
virialized 
system, this improvement of the TFR implies that we can use the values of \SK\ to measure the 
dynamical masses of virialized systems (we investigate this in more detail in
\S 5 and in the appendix).   One potentially issue is that we are using the velocity dispersion as measured
from the kinematics of the gas rather than the stars.  However, as has been shown by
e.g., Kobulnicky \& Gebhardt (2000) there is a strong correlation between these two quantities.
Furthermore, as there is a strong, low scatter, correlation between S$_{0.5}$ and
stellar mass (e.g., Kassin et al. 2007; \S 5.1.2) this suggests that the kinematics of
the gas is tracing the underlying dynamical properties. Thus, as 
non-virialized systems obey the same TFR, we can likely 
derive their dynamical masses in a similar way.

The dynamical mass is then calculated using the \SK\ values for our primary and secondary samples as:
        
        \begin{equation}
          \mbox{M}_{\rm{dyn}}(r_{e}) = \frac{S_{0.5}^{2}r_{e}}{G}.
          \label{eqn:Mvir}
        \end{equation}

\noindent Furthermore, we use the $S_{0.5}$ index to measure the effective velocity 
dispersion which differs slightly from 
previous work.  For systems that are pure spheroids without any or little 
rotation, the value of 
$S_{0.5}$ is nearly identical, or exactly identical, to using the velocity 
dispersion.  For systems 
that are pure rotational velocity systems, the value of the dynamical mass is as if it were being 
measured using simple kinematics and measuring the mass content within a fraction of a
scale length.

 This closely relates to other measures of dynamical masses used in the past 
(e.g., Conselice et al. 2005; Treu et al. 2005), but in effect uses both the rotational and 
internal kinematics, while previous examinations have used one or the other. If we were to calculate the total amount of mass in galaxy then we would use a coefficient in front of eq. (2) - with typical values between 2-5 (e.g., Binney \& Tremaine 1987).  However, as described above, we are using this definition of the dynamical mass in its basic form without any interpretation of what mass it is measuring as this can vary with the type of galaxy. Ultimately, we are interested in relating this quantity to the total or halo masses of galaxies.

We also calculate at certain times in this paper what we call the dynamical $\beta$ mass which is also calculated using the \SK\ values as:
        
        \begin{equation}
          \mbox{M}_{\beta, \rm{dyn}}(r_{e}) = \frac{\beta (n) \times S_{0.5}^{2}r_{e}}{G}.
          \label{eqn:Mvir}
        \end{equation}

\noindent where the value of $\beta$ depends on the Sersic index, $n$ in the following way (Cappellari et al. 2006):

\begin{equation}
\beta(n) = 8.87 - 0.831\times n + 0.0241\times n^{2}.
\end{equation}

\noindent Where we have used the fitted Sersic profiles for our galaxies based on surface brightness fits (e.g., Trujillo et al. 2007).  Note that although Sersic fits do not represent the total light distribution in merging galaxies, they are still able to represent the broad light profiles of these systems.   The value of $\beta$ varies between $\beta \sim 6$ for n = 4 and $\beta \sim 8.1$ for galaxies with Sersic indices $n = 1$.
We explore how changes to this index would alter our results, but find in general that using a canonical value of $\beta = 5$ would not change our results in any significant way.
 This closely relates to other measures of dynamical masses used in the past, 
but in effect uses both the rotational and internal kinematics, while previous examinations have used one or the other.

        \begin{table}
          \begin{tabular}{l l l l}
            \hline
            Mass Type & Symbol & Equation & Section \\
            \hline
Stellar  & M$_{*}$ & Derived & \S 3.1 \\
Dynamical & M$_{\rm dyn}$ & $\frac{{\rm S}_{0.5}^{2} r_{e}}{G}$ & \S 3.2 \\
$\beta$, Dynamical & M$_{\rm \beta, dyn}$ & $\frac{\beta {\rm S}_{0.5}^{2} r_{e}}{G}$ & \S 3.2 \\
Halo & M$_{\rm halo}$ & $a \times ({\rm log M_{*}}^{b} + c)$ & \S 4.1 \\
Halo Virial & M$_{\rm halo,vir}$ & $\frac{{\rm V}_{\rm halo}^{2} r_{\rm halo}}{{\rm G}}$ & \S 4.2 \\
Circular & M$_{\rm circ}$ & $\frac{{\rm V}_{\rm circ}^{2} r_{1/2}}{{\rm G}}$ & \S 4.3.2 \\
Sigma ($\sigma$) & M$_{\rm sig}$ & $\frac{\sigma^{2} r_{\rm e}}{{\rm G}}$ & \S 4.3.3 \\
            \hline
          \end{tabular}
          \caption{The definitions of various masses used throughout this paper.  The section in which they are defined and explained are also listed.   Note that our definition of some of these terms, especially dynamical mass, differs from previous works (see text). }
          \label{tab:CAS}
        \end{table}

\section{Measuring Total Halo Masses}

        \label{sec:haloMass}

In this section we describe our derived methods for measuring the halo masses of 
galaxies from observations and theory/models, namely the stellar mass and/or the internal velocities and sizes of 
individual galaxies, and through abundance matching.  Ultimately we use theoretical models
which relate observables to halo mass.  We also discuss how similar different mass measures are when compared with each other.

Galaxies reside within dark halos with physical extents which are much larger than 
their visual radius (e.g., Navarro, Frenk \& White 1996; Persic, Salucci \& Stel 1996; 
Capellari 2006; Kratsov 2013), and measuring their total dark halo masses (hereafter referred to 
simply as the halo mass) of galaxies is a non-trivial exercise.  One can do this to some 
degree in the local universe by measuring HI velocity rotation curves to a large radius, but 
there are very few distant galaxies that have such measurements.   Likewise, it is possible 
to use strong gravitational lensing to measure galaxy masses, but the examples of this are 
rare, and it is not clear if these galaxies have especially concentrated light profiles, 
and thus would not be representative of galaxies in general.  Furthermore lensing can 
only be used 
in a very small number of systems, but we are interested in a more generalized method 
for finding the 
halo masses of individual galaxies.

        \begin{figure*}
          \centering
	\vspace{-7cm}
          \hspace{-1cm} \includegraphics[angle=0, width=19cm]{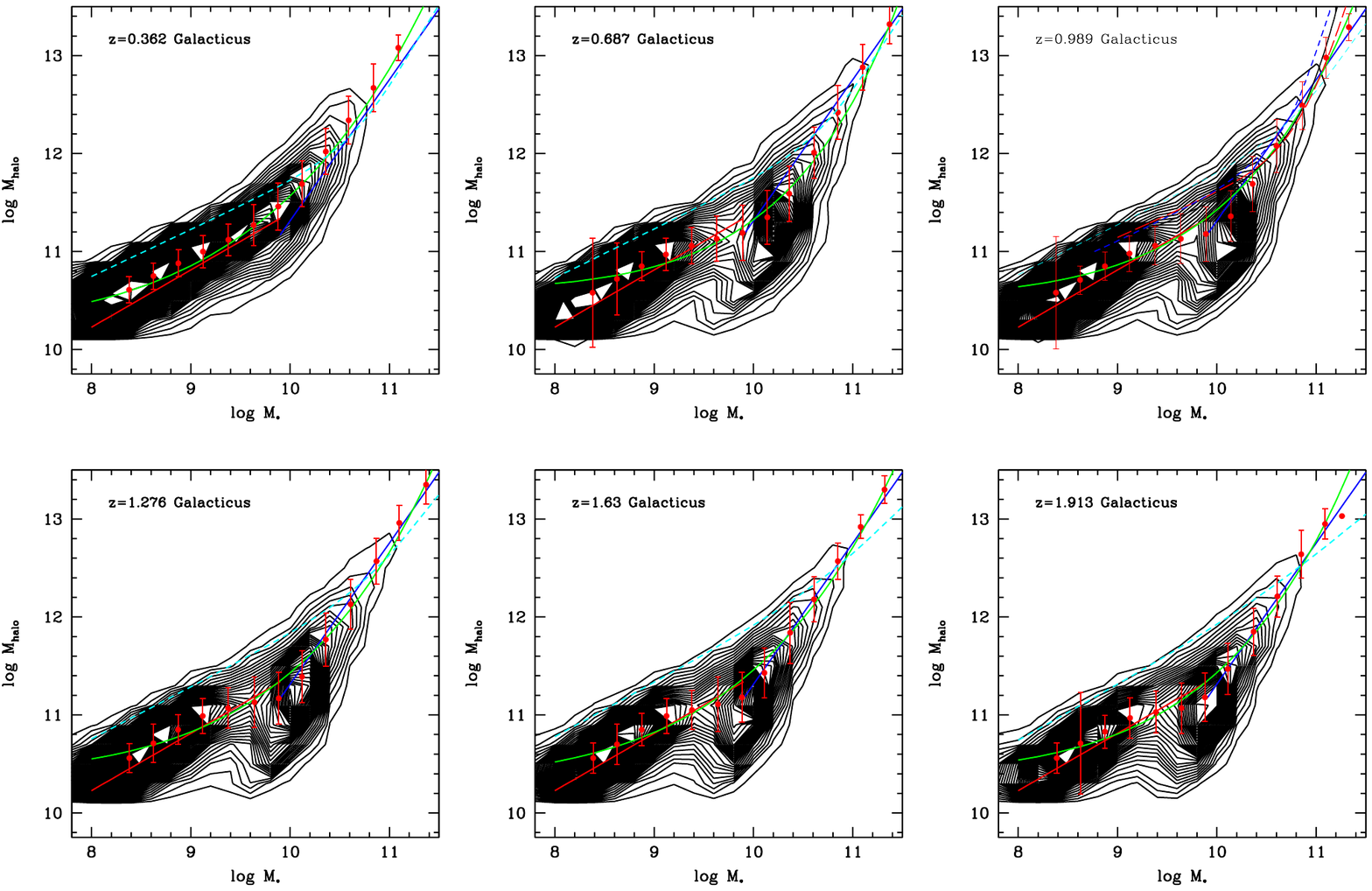}
\vspace{-7.5cm}
          \caption{The relation between the stellar mass and the halo mass for galaxies within 
the Galacticus simulation results from Benson et al. (2012).  Show here as the contour plot are 
the results for these models.  The green line shows the best fit to the Galacticus 
relations, while the red points are the averages of the models at different stellar masses, and their 1 $\sigma$ dispersions.  The red and blue lines show the fit when using linear relations, with a break at log M$_{*} \sim 9.6$.  Note that there is very little 
evolution in terms of the stellar and halo masses as a function of redshift up to $z \sim 2$.   We compare our results with empirical measurements from clustering and lensing in the $z = 0.989$ panel.  
The black curved line is the relationship between stellar and halo mass from the weak lensing analysis of van Uitert et al (2016), the red curved line is from the lensing results of Leauthaud et al. (2012), and the blue dashed line is from Moster et al. (2013).  The SHAM model of Rodr{\'{\i}}guez-Puebla et al.(2017) is shown as a dashed cyan line. }
 
        \end{figure*}
       
We therefore must resort to other methods that utilize empirical and model calculations for how halo mass 
relates to observable properties.   This includes using rotation curves, internal 
velocity dispersions, galaxy abundance matching and clustering - which all can give measures of average
halo masses for a selected population (for example Conselice et al. 2005; Weinmann et al. 2006; Epinat et al. 2009; 
Foucaud et al. 2010; Wake et al. 2011; Leauthaud et al. 2012).

Since measuring the halo mass is an important issue, we describe in some detail how we estimate this 
quantity, and describe the uncertainties associated with these inferences.  We first discuss how 
the stellar and halo masses relate for galaxies in simulations as a new approach for obtaining the halo 
masses of systems.  We also check to see how similar the results of this method are in comparison with 
other methods of obtaining halo masses.

We first describe a model method for measuring the halo mass through relating the observable stellar 
mass to the halo mass based on simulation output.  We then describe a galaxy abundance matching
 technique which relates halo mass directly to the observed galaxy stellar mass function, using our
own measured stellar masses. We finally 
describe how these measures relate to the ratio of stellar and halo masses derived through clustering. 
 By using, and comparing, several methods to obtain the halo mass we quantify how well any one technique may be 
doing, and quantify the systematics that may be present.  

\subsection{Halo Masses from Stellar Masses using Models}
        
We present a new way to calculate halo masses from observations by using a 
semi-empirical/semi-analytical method similar in spirit  to that used by 
Conselice et al. (2005) to calculate the \Mtot\ values 
from \Mvir.   Previously, Conselice et al. (2005) fit the ratio of 
\Mtot\ to \Mdyn\ as a function 
of \Mdyn\ from semi-analytical models where both \Mdyn\ and \Mtot\ are predicted.  We investigate
this in two ways - by measuring halo masses from stellar masses and later by using kinematics
(\S 4.3).

We thus investigate the relationship between the values of M$_{\rm *}$, as defined  in 
\S 2, to the halo mass as found in the semi-analytic models of 
De Lucia \& 
Blaizot (2007) and the Galacticus simulation from Benson et al. (2014).   These are both semi-analytical models that use similar methods and merger trees, but the details of how the astrophysics is included differ.  The details of Galacticus can be found in Benson et al. (2014), but we summarize them here for completeness.  Galacticus is similar to other semi-analytical codes, such as the one by De Luca et al. but is more flexible and adaptable to different input conditions and parameters.   In Galacticus star formation occurs when gas is accreted from the intergalactic medium at a rate which is proportional to the growth rate of the halo.  Both models have standard prescriptions for feedback, star formation, and the cooling of gas. The Galacticus input is a simplified model of galaxy formation designed to match observations in the local universe, including the galaxy stellar mass function.     We carry this out for all stellar masses and thus investigate 
how this relationship changes as a function of the input stellar mass.

The halo masses in the simulations we use are virial masses, defined as where the density is 200 times
the critical density, such that 
M$_{\rm halo} = M_{200} = (4 \pi/3)\times 200 \rho_{\rm critical}(z)R_{200}^{3}$. Specifically,
in Galacticus the virial mass is defined in the same manner as used in Bryan \& Norman (1998),
where the density contrast is utilized,  implying that it is 
equal to the virial
density assuming a spherical collapse model.  To address this we first use the results of the Galacticus semi-analytical model (Benson et al. 2012) 
to determine how the relationship between stellar mass and halo mass evolves with time.
 
We show how the stellar mass changes with halo mass in Galacticus from 
$z \sim 0.3$ to $z \sim 2$ in Figure~2. As 
can be seen, there is a change in slope in the correlation between the 
stellar and halo mass at around M$_{*} \sim 10^{10}$ \msol.  We find that
we can fit this relation as either two straight lines (see Appendix) 
with a break at this 
stellar mass, or as a power-law in log-space as given by:

\begin{equation}
{\rm log M_{\rm halo}} = a \times ({\rm log M_{*}})^{\rm b} + c,
\end{equation}

\noindent where the best fits for these values are shown in Figure~2 as the green solid
lines.    An important 
component of this relation is not just the best fit, but also the scatter in the values.  The 
red error bars on Figure~2 show the scatter in the relation between the stellar mass and the 
halo mass at these various redshifts.   We further discuss this scatter and how it can be 
minimized in the next section. We also compare with other results relating the stellar mass to halo mass relation, including the relations published in Rodr{\'{\i}}guez-Puebla et al.(2017) who use subhalo abundance matching (SHAM) to retrieve their relations between stellar and halo masses.  We find
some differences with this model, particularly at the lowest masses, which is likely an indication
that there are subhalo masses being confused with larger host halo masses.

        \begin{table}
          \begin{tabular}{c c c c}
         
            \hline
            Redshift & $a$  & $b$ & $c$  \\
            \hline
              0.40  & 1.39$\pm0.23 \times 10^{-7}$ & 6.99$\pm$0.5 & 10.2$\pm$0.1  \\
              0.70  & 3.89$\pm0.63 \times 10^{-11}$& 10.3$\pm$1.1 & 10.6$\pm$0.1 \\
              1.00  & 4.09$\pm0.63 \times 10^{-8}$& 9.3$\pm$1.2 & 10.5$\pm$0.1   \\
              1.30  & 7.61$\pm0.16 \times 10^{-7}$& 8.1$\pm$1.1 & 10.4$\pm$0.2  \\
              1.60  & 1.89$\pm0.71 \times 10^{-7}$& 7.8$\pm$1.2 & 10.3$\pm$0.2   \\
              1.90  & 2.08$\pm0.87 \times 10^{-8}$& 8.7$\pm$1.1 & 10.4$\pm$0.1  \\
            \hline
          \end{tabular}
        \caption{The fitted values for eq. 5 which relates the stellar mass and halo
mass for galaxies from the Galacticus simulation. (\S 4.1). }
        \label{tab:stellarhalo}
        \end{table}

\noindent We calculate that the scatter in this fit varies between 0.13 \,dex for systems at log M$_{*} = 9$ to a scatter of 0.37 at around log M$_{*} \sim 11$ in both the Galacticus and the Millennium simulations.  Table~3 shows the coefficients of this relation at different redshifts.  

This implies that by just using eq. (5) one can, if these models are correct, obtain measurements of the halo mass from the stellar mass within a factor of 2.5 or better, which is slightly larger than the typical uncertainty on stellar mass measurements themselves (e.g., Mortlock et al. 2011, 2015).  As we discuss in \S 4.2, significant physics is present within this scatter, with a difference seen between red and blue central galaxies at the higher mass end of this range (e.g., Rodriguez-Puebla et al. 2015).  Furthermore,  as we also later find, Rodriguez-Puebla et al. (2014) report  that the scatter increases as higher masses, largely due to a higher differential in formation histories between red and blue centrals.

There are several uncertainties in measuring the halo mass via this method, some of which are discussed in Conselice et al. (2005).  One major issue is halo occupation, such that more than one galaxy is in a single halo (e.g., Berlind et al. 2003).  However, our relation is derived for individual sub-halo and galaxy masses, and not the overall larger halo in which it may exist. This limits our ability to compare with results that measure the halo mass of a given sample, as is done through e.g., clustering or lensing.  This is particularly the case for lower mass systems that are likely within more massive overall halos.   Furthermore, these galaxy halo masses are only expected to be accurate to a factor of a few, given the scatter in the relation.    There are however potential systematics that we investigate by comparing our results to those which are obtained through other methods.

\subsection{Third Parameter Effect}

One of the major questions in trying to understand the relation between halo mass and galaxy stellar mass is whether there is a third parameter which affects the mapping between these two masses.  There is some evidence that this is indeed the case, as the relation between halo and stellar mass has a dependence on colour that is fairly strong (e.g., Hearin et al. 2013; Rodriguez-Puebla et al. 2015).  In fact, there are various ways in which the relation between the halo mass and stellar mass can be improved by investigating the detailed dependence on colour, as a proxy for star formation history, and then explicitly accounting for it (e.g., Hearin et al. 2013; Hearin et al. al 2015; Rodriguez-Puebla et al. 2015).  

Colour is the easiest parameter beyond luminosity to measure for a galaxy, yet colour itself however depends on several factors, the most important of which is the time-scale in which the galaxy is quenched.     The fact that a galaxy's colour is  part of the extra parameterization of matching galaxies to halos can be seen in galactic conformity, whereby central and satellite galaxies within a halo have similar colours (i.e., both are either red or blue) (e.g,. Kauffmann et al. 2013).    This relation holds even up to higher redshifts (Hartley et al. 2015), and therefore it is likely a fundamental way in which galaxy halo and stellar mass relate.   However, colour is likely only an effect of a more fundamental underlying 3rd parameter between the the halo and stellar masses of galaxies, and we investigate this problem more broadly using our simulation results.

We investigate this third parameter effect using the Galacticus simulation output.  In detail, we investigate how the scatter in the stellar mass to halo mass relation changes when considering other parameters. 

Namely we consider the merger history, the concentration of the virialized halo ($C_{\rm vir}$), the time when half of the mass of the halo was formed in Gyr since the Big-Bang (t$_{\rm form}$), a combination of internal velocities of the halo $V_{\rm halo,max}$ and  size at R$_{200}$ ($R_{\rm halo}$) to measure a virial mass, as well as the dynamical mass as defined in \S 3.2. We convert the rotation and size into a virialized halo mass defined by:

\begin{equation}
{\rm M_{halo,vir} = \frac{V_{\rm halo}^{2} \times R_{\rm halo}}{\rm G}}
\end{equation}

\noindent which is related to the dynamical mass, but is a simplier formulation.  Overall, what we find is that all of these values can act as a third parameter when investigating the relation between the stellar and halo mass.  

To investigate this quantitatively, we consider a linear parameterization such that,

\begin{equation}
{\rm log M_{halo}} = {\rm a} \times {\rm log M_{*}} + {\rm b} \times {\rm M_{halo,vir}} + c
\end{equation}

\begin{equation}
{\rm log M_{halo}} = {\rm a} \times {\rm log M_{*}} + {\rm b} \times {\rm C_{vir}} + c
\end{equation}

\begin{equation}
{\rm log M_{halo}} = {\rm a} \times {\rm log M_{*}} + {\rm b} \times {\rm t_{form}} + c.
\end{equation}

\begin{equation}
{\rm log M_{halo}} = {\rm a} \times {\rm log M_{*}} + {\rm b} \times {\rm M_{dyn}} + c.
\end{equation}

\noindent We show the resulting fitted values in Table~4-7 for the best $\chi^{2}$ fit using these three features as a third parameter.  We find that at higher masses, the most effective third parameter for reducing the scatter in the fit between the stellar and halo mass is the time-scale in which the galaxy halo is assembled and the dynamical mass (Figure~3).  Within the Galacticus simulation, this time-scale is defined as
the age of the universe when half of the halo mass is assembled.  This correlates well, but with
some scatter, with the time of the last major merger within this simulation.

We can also see from Figure~3 that there is a reduction in the scatter when considering the other possible third parameters, including the virial mass, dynamical mass, and the halo concentration.  These are all related, as the halo concentration correlates to the formation history of the halo, as well as to its colour (e.g., Wechsler et al. 2002; Hearin \& Watson 2013).    This is not quite as simple in the Galacticus simulation, however, where the concentration for main halos is determined by the halo mass, but it should be present for the satellite galaxies.  This is because for satellites the values of  C 
and $t_{\rm form}$ will correlate, as the value of C  is determined from when a satellite falls into a larger halo, and
by definition $t_{\rm form} < t_{\rm infall}$.  However for the bulk of the galaxies this is not the case, which is one reason why concentration is not as good of a third parameter as the other variables.  Therefore these four parameters appear to be the most suitable third parameters that allows us to use the relation between the stellar and halo mass to a higher accuracy.  

If we take the example of the time-scale of formation, as given by the t$_{\rm form}$ parameter, then the sense of this correlation is such that at a given stellar mass, the derived halo mass is lower if formation occurred at an earlier time.  This implies that galaxies which formed earlier in the universe would have a lower value of M$_{*}$/M$_{\rm halo}$ than galaxies which formed later.  We later test this idea in \S 5.1.    Note also from Figure~3 that the dynamical mass produces the lowest scatter between M$_{*}$ and M$_{\rm halo}$.  This
correlation is such that the dynamical mass is higher for higher mass halos at  given fixed stellar mass.

       \begin{figure}
          \centering
	\vspace{-1cm}

          \includegraphics[angle=0, width=8.5cm]{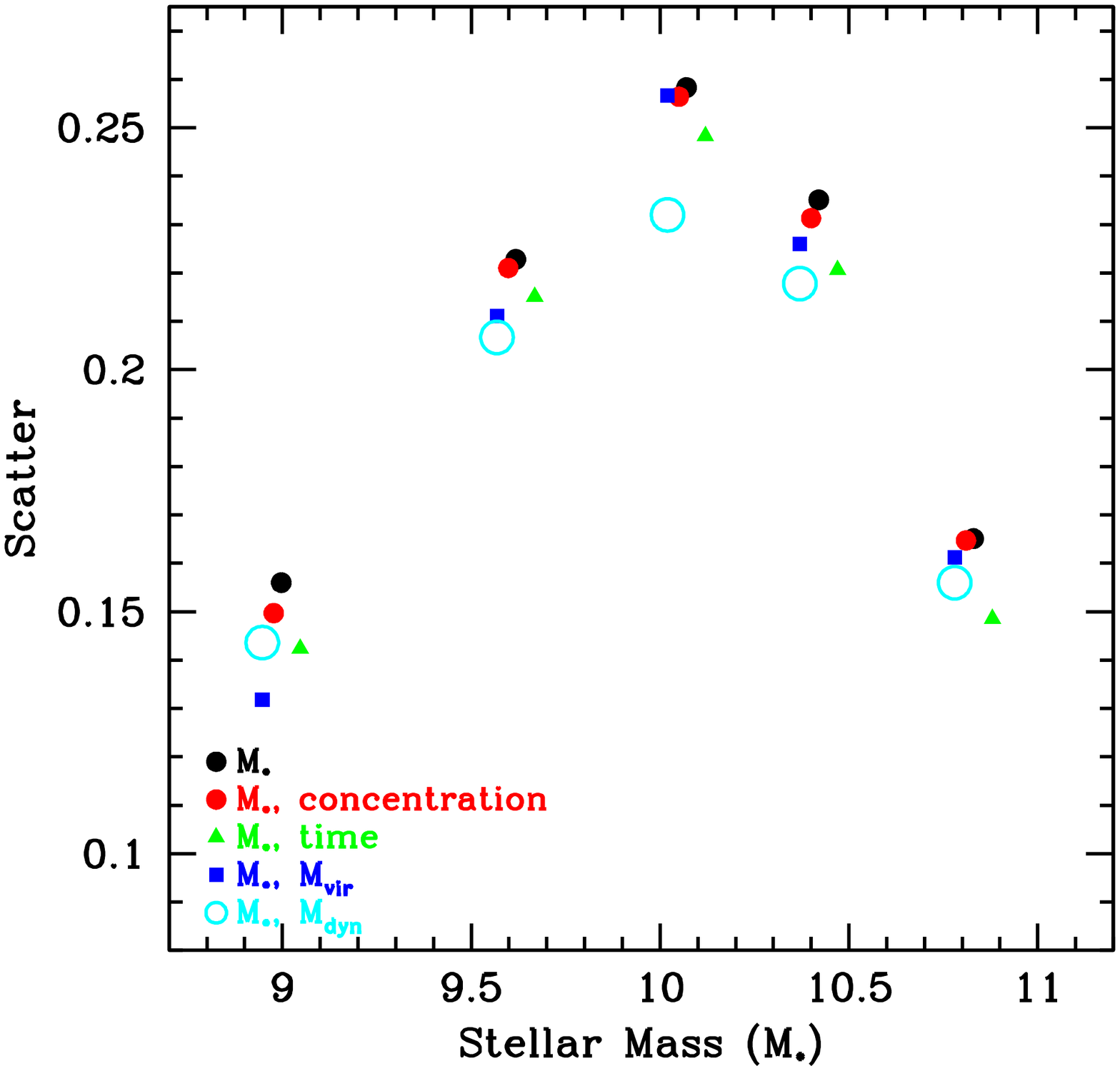}
	\vspace{-2cm}

          \caption{The scatter in the relation between the measured halo mass from the Galacticus simulation output and the predicted values based on various parameterizations using a third parameter (\S 4.2).  The solid black points show the scatter when just considering the relation between stellar mass of the galaxy and the halo mass.  The solid red symbols show the relation when considering stellar mass and the halo concentration in a 3rd parameter fit, while green triangle shows this relation for stellar mass and time of the halo formation, the blue boxes the relation scatter when examining the fit between the stellar mass and the virial mass, and the open cyan is for the
dynamical mass.   }
          \label{fig:S05MbaryTFR}
        \end{figure}

       \begin{figure*}
          \centering
\vspace{-2cm}

          \includegraphics[angle=-90, width=19cm]{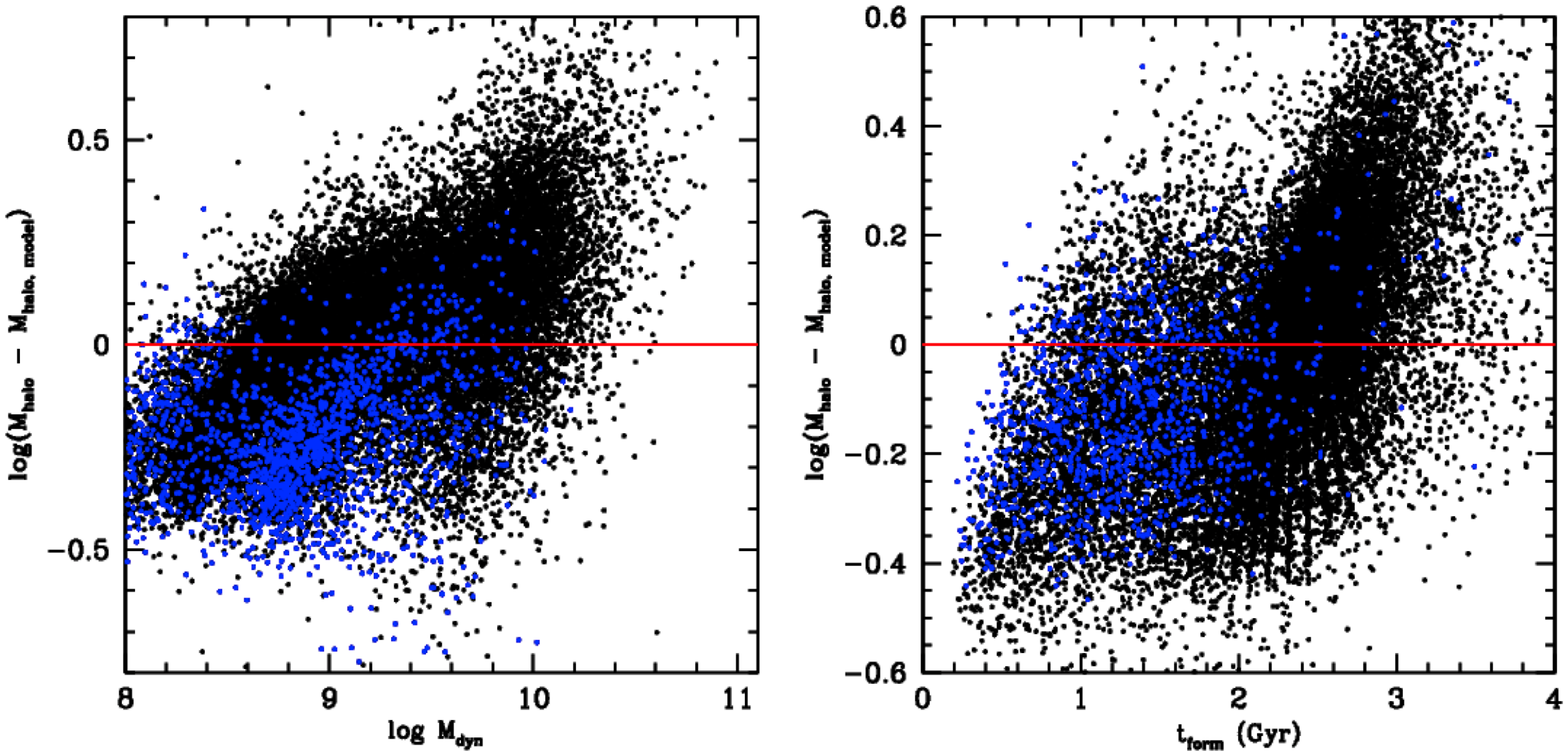}
\vspace{-3cm}
          \caption{Relation between the halo mass of galaxies in the Galacticus simulation at $z = 1$ minus the model fit mass based on the stellar mass (eq. 5).  In the left panel we show the correlation of this property with the dynamical mass, while on the right panel we show it vs. the time of formation.  The difference is such that systems which are overfitted by the average best fit have an earlier time of formation and a lower dynamical mass.  Whereas those systems that are more massive than the best-fit formed later in the simulation and have a higher dynamical mass.  The blue points in the left panel show systems which have a time of formation $< 1$ Gyr, and on the right the blue points are those systems which have a ratio of
dynamical to halo mass $< 0.005$. }
          \label{fig:S05MbaryTFR}
        \end{figure*}

        \begin{figure*}
          \centering
\vspace{-4cm}
          \hspace{-1cm} \includegraphics[angle=-90, width=19cm]{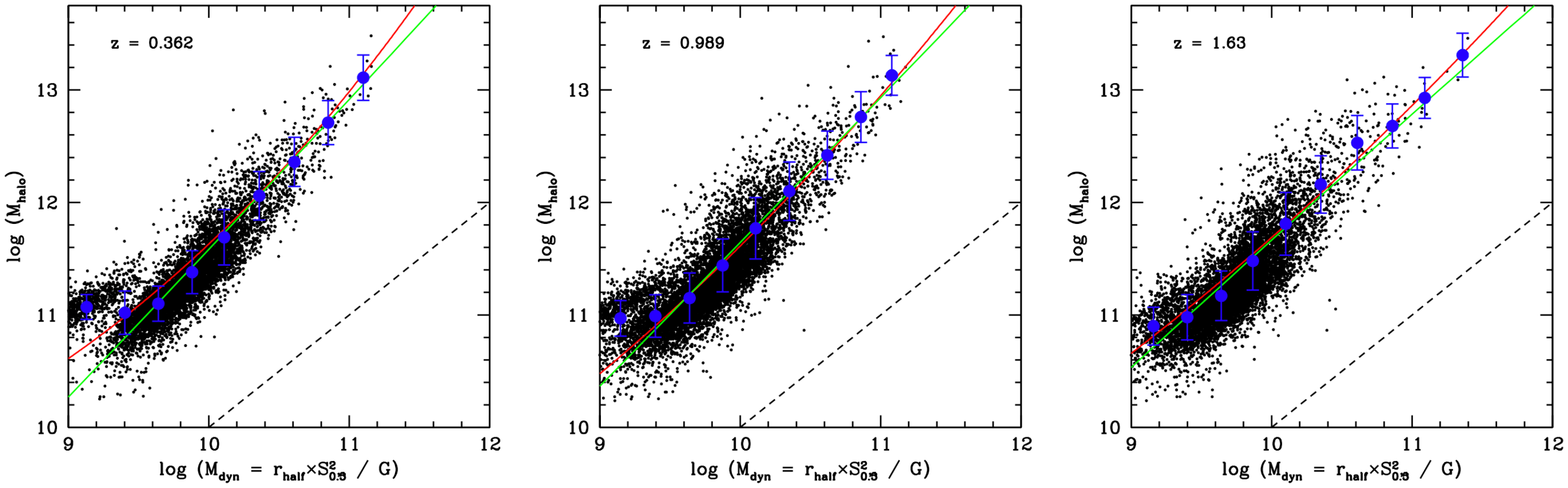}
\vspace{-4cm}

          \caption{The relation between the dynamical mass, as measured with the S$_{0.5}$ parameter (\S 3.2), and the halo mass from the Galacticus simulation.  The dashed line shows where the 1:1 ratio for these parameters would be, while the red line shows the best fitting relationship between the halo and dynamical mass. The green line shows the Moster et al. (2010) fitting formalism to this relation (see Appendix).} 
          \label{fig:s05}
        \end{figure*}

To further demonstrate this, we show that the difference in the halo mass and the best fit model mass based on just the stellar mass (eq. 5) has residuals that correlate with both the dynamical mass and the time-scale of formation (Figure~4).  This is such that the scatter in the relationship correlates with each parameter whereby those masses that are overfit (negative values) formed earlier and have lower dynamical masses, and those which are underfit (positive values) have later formation times and higher dynamical masses.

        \begin{figure*}
          \centering
\vspace{-4cm}

          \hspace{-1cm} \includegraphics[angle=-90, width=19cm]{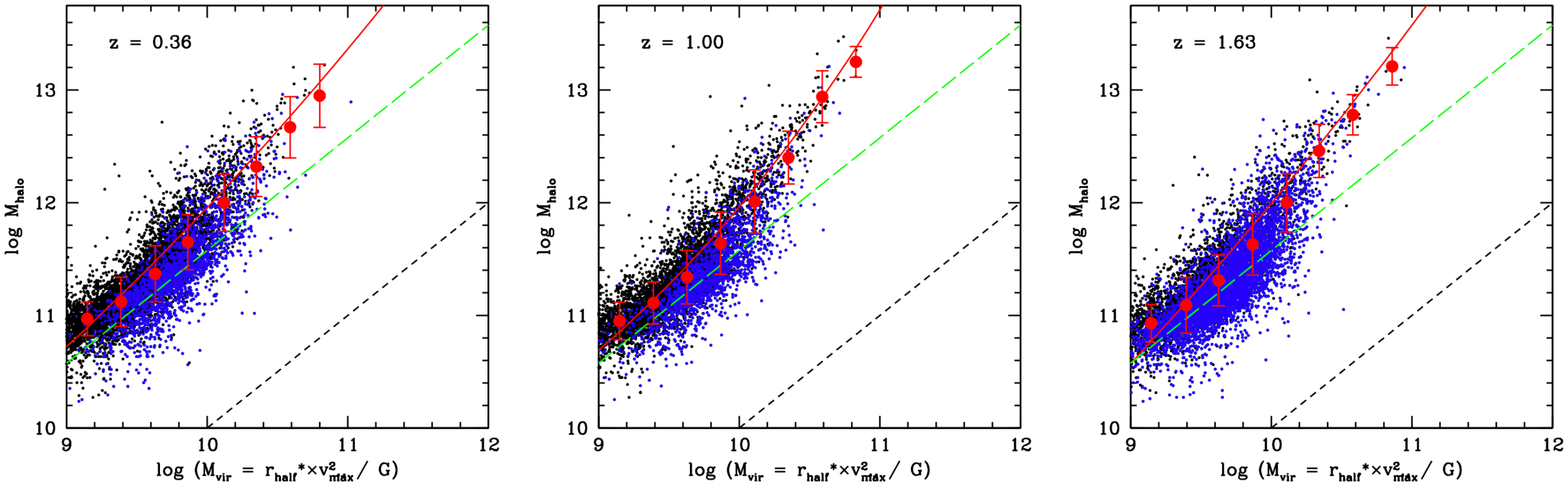}
\vspace{-4cm}
 
         \caption{The relation between the circular velocity mass (\S 4.3.2), as measured with the V$_{\rm max}$ parameter and the halo mass from the Galacticus simulation.   This is a typical method for finding the masses of disk like or rotating galaxies.  We show as the blue points rotation dominated systems which have a V$_{\rm max} / \sigma > 1$. The dashed line shows where the 1:1 ratio for these parameters would be, while the red line shows the best fitting relationship between the halo and virial masses.  The green long dashed line shows the value of the halo mass derived from the circular mass when using the analytical relation of half-light radius to the virial radius (eq. 12).  } 
          \label{fig:s05}
        \end{figure*}

        \begin{figure*}
          \centering
\vspace{-2cm}

          \includegraphics[angle=-90, width=16.5cm]{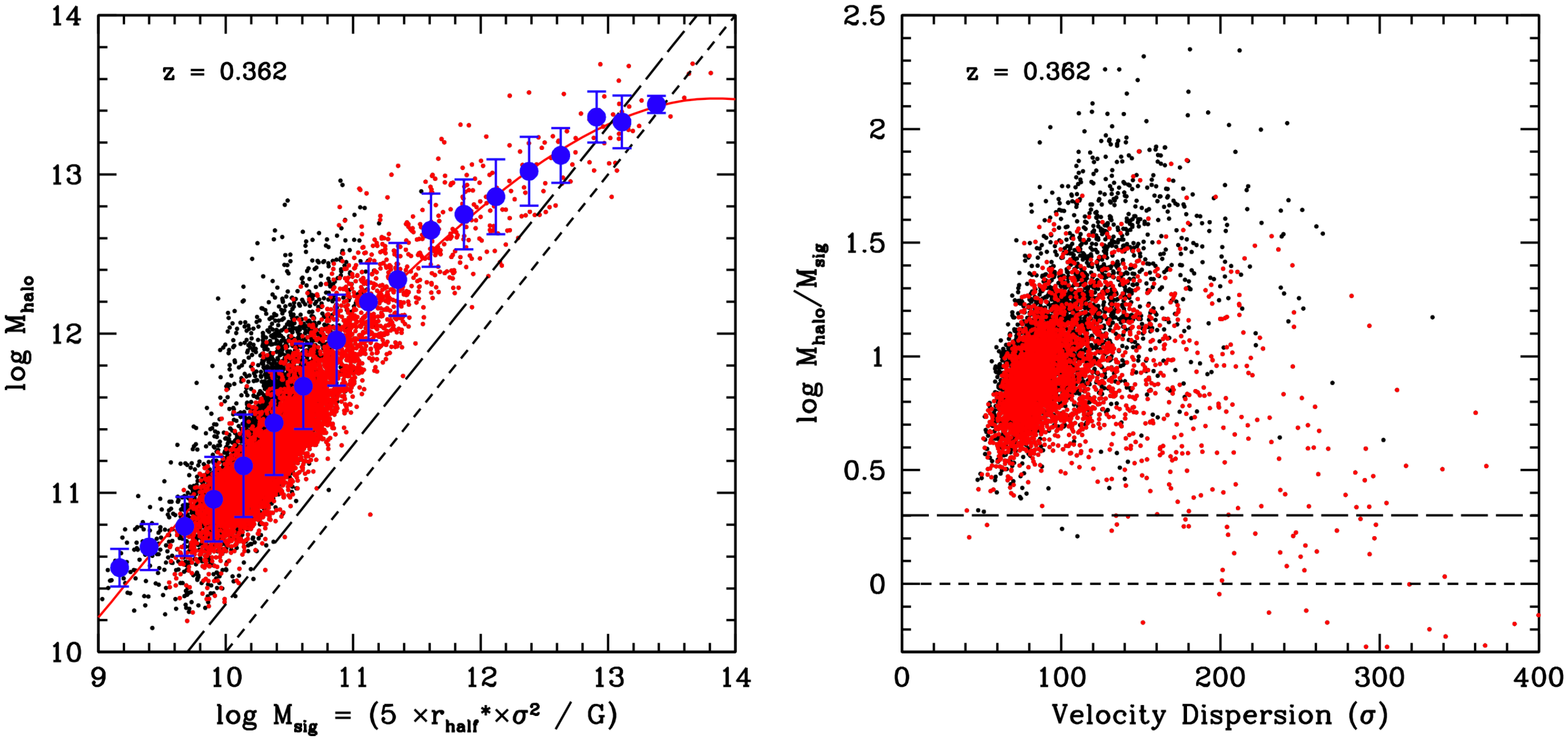}
\vspace{-1.75cm}

          \caption{Figures showing the singular use of the velocity dispersion to retrieve the halo 
masses of
galaxies.  The left hand panel shows the relation between the sigma mass (see text) and the halo
mass from the Galacticus models.  The blue points show the averages of all galaxies, while the red
dots show galaxies with a ratio of rotational velocity and velocity dispersion such
that (V$_{max} / \sigma) < 1$.  The solid red line shows the best fit between the sigma mass and
the halo mass using the formalism from Moster et al. (2010).  The short dashed lines show the 1:1 ratio,
while the longer dashed line show the ratio of 2:1 for underestimating the halo mass by a factor of
two based on the sigma mass.   The right hand panel shows the difference between the halo mass and
the sigma mass for galaxies at z = 0.362.  The two horizontal lines show the 1:1 and 2:1 ratio as
described for the left hand panel.} 
          \label{fig:s05}
        \end{figure*}

        \begin{figure}
          \centering
\vspace{-2cm}

          \includegraphics[angle=0, width=8.5cm]{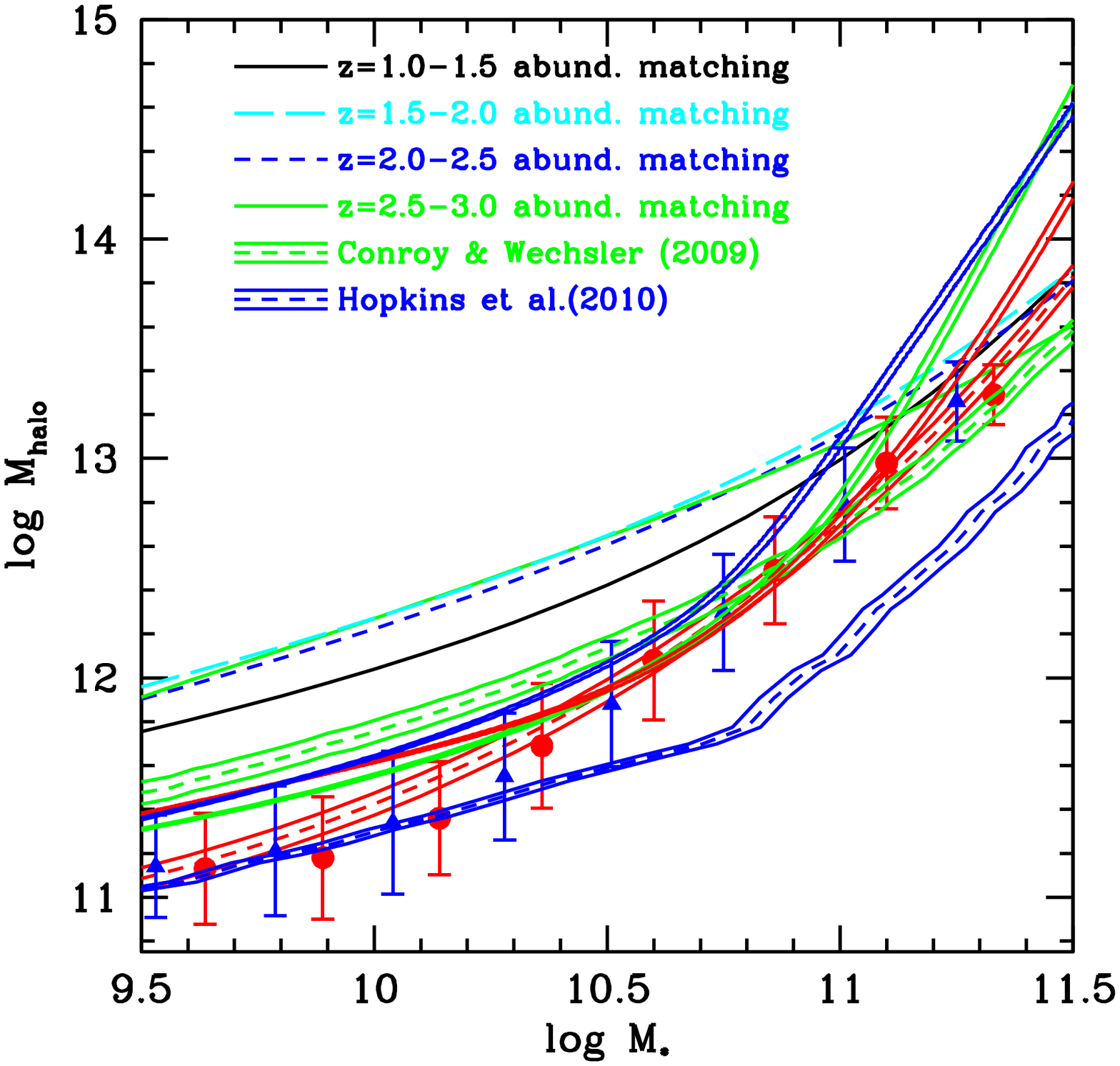}
\vspace{-2cm}

          \caption{The relation between stellar mass and halo mass as derived through abundance matching and weak lensing.  Shown are four lines from this method representing the relationship between these two quantities from $z = 1$ to $z = 3$, as shown by the curved single lines.  The points (red for $z \sim 1$ and blue for $z \sim 0.4$) show the relationship between the stellar and halo masses as derived through the Galacticus simulation (\S 4.1).  The dashed red line surrounded by two solid lines shows the best fitting relationship for this simulation at $z \sim 1$.     The double solid green curved line is the relationship between stellar and halo mass from weak lensing measures from van Uitert et al (2016), the solid red double line is from the lensing results of Leauthaud et al. (2012), and the blue solid line is from Moster et al. (2013).   We also show the abundance matching stellar and halo masses from Hopkins et al. (2010) and Conroy and Wechsler (2009).}
          \label{fig:abundance}
        \end{figure}

       \begin{figure}
          \centering
\vspace{-2cm}

          \includegraphics[angle=0, width=8.5cm]{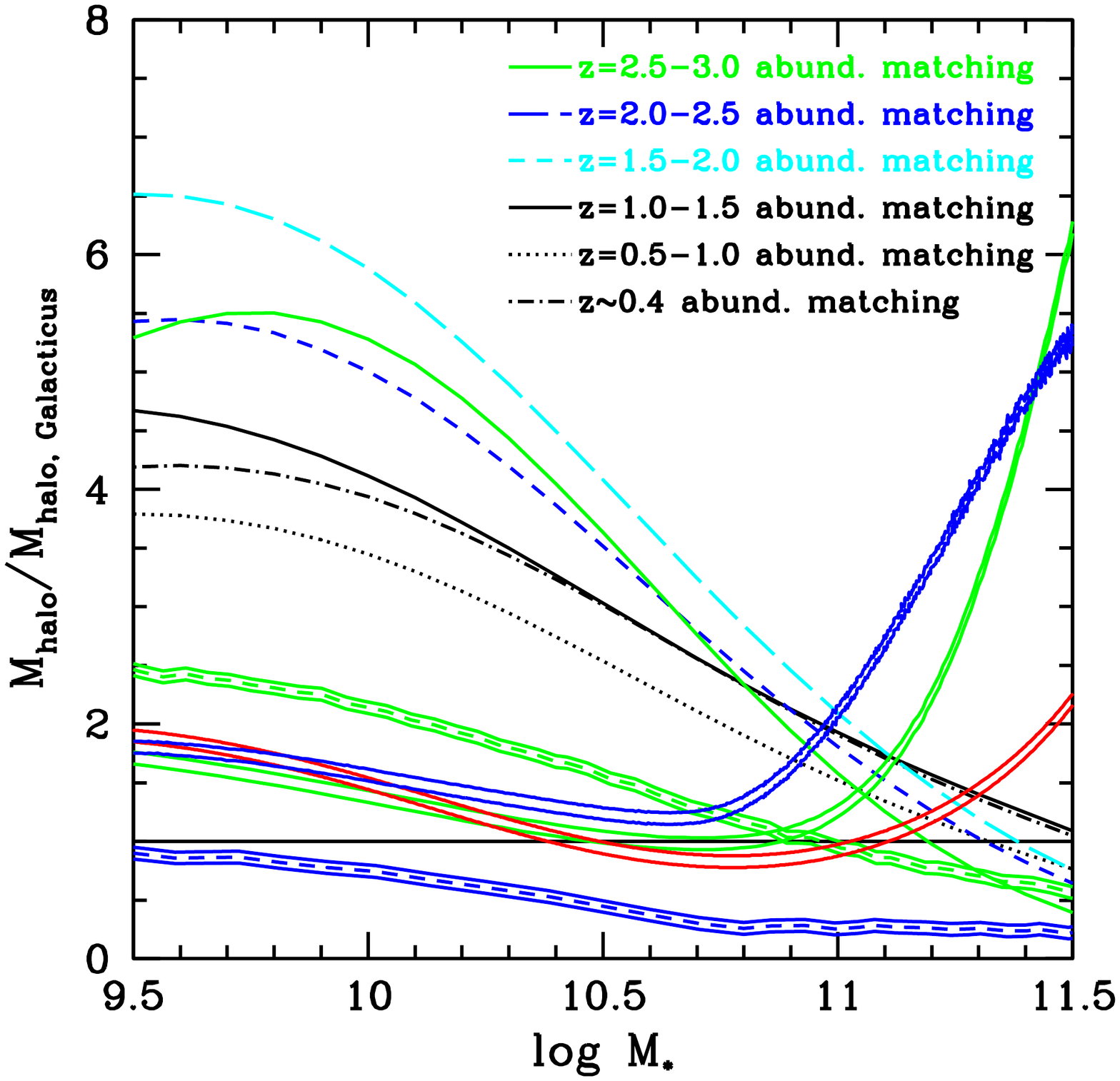}
\vspace{-2cm}

          \caption{Plot showing the ratio between the halo masses measured with abundance matching (and weak lensing)
at a given stellar mass from log$_{*}$ = 9.5 to 11.5 compared to the predictions from
the Galacticus simulation (\S 4.1).  The key shows the similar lines as in Figure~8.    The double solid green curved line is the relationship between stellar and halo mass from weak lensing measures from van Uitert et al (2016), the red double line is from the lensing results of Leauthaud et al. (2012), and the blue double line is from Moster et al. (2013).  As in Figure~8 the dashed line with solid outlines is from Hopkins et al. (2010), and the green dashed line with solid outlines is from 
Conroy and Wechsler (2009). The horizontal line shows where our measurement of halo mass
from Galacticus is identical to these other methods.}
          \label{fig:S05MbaryTFR}
        \end{figure}

        \begin{table}
         
            \hspace{2.0cm} log M$_{*} < 9.9$ \hspace{1.0cm} log M$_{*} >  9.9$  \\
          \begin{tabular}{c c c c c c c}
            
            \hline
            Redshift & $a$  & $b$ & $c$ & $a$ & $b$ & $c$ \\
            \hline
              0.4  & 0.12 & 0.58 & 4.46 & 2.09 & -0.82 & -1.57 \\
              0.7  & 0.12 & 0.56 & 4.60 & 2.01 & -0.68 & -2.02 \\
              1.0  & 0.15 & 0.51 & 4.82 & 1.79 & -0.40 & -2.52 \\
              1.3  & 0.18 & 0.47 & 4.98 & 1.63 & -0.24 & -2.51 \\
              1.6  & 0.21 & 0.43 & 5.10 & 1.49 & -0.06 & -2.88 \\
              1.9  & 0.22 & 0.41 & 5.12 & 1.36 & 0.10 & -3.15 \\
            \hline
          \end{tabular}
        \caption{The fitted values of a, b and c for the relation between stellar mass, halo mass, and $M_{\rm halo,vir}$ as found through the Galacticus simulation. These values are used in equation (7).  }
        \label{tab:stellarhalo}
        \end{table}

        \begin{table}
            \hspace{2.2cm} log M$_{*} < 9.9$ \hspace{1.5cm} log M$_{*} >  9.9$  \\
          \begin{tabular}{c c c c c c c}
            
            \hline
            Redshift & $a$  & $b$ & $c$ & $a$ & $b$ & $c$ \\
            \hline
              0.4  & 0.53 & 0.04 & 5.78 & 1.25 & -0.05 & -0.68 \\
              0.7  & 0.52 & 0.04 & 5.82 & 1.37 & -0.02 & -2.08 \\
              1.0  & 0.52 & 0.05 & 5.82 & 1.47 & 0.02 & -3.46 \\
              1.3  & 0.51 & 0.07 & 5.82 & 1.49 & 0.06 & -3.89 \\
              1.6  & 0.51 & 0.10 & 5.72 & 1.51 & 0.12 & -4.35 \\
              1.9  & 0.50 & 0.16 & 5.53 & 1.52 & 0.16 & -4.55 \\
            \hline
          \end{tabular}
        \caption{The fitted values of a, b and c for the relation between stellar mass, halo mass, and $C_{\rm vir}$ as found through the Galacticus simulation. These values are used in equation (8).  }
        \label{tab:stellarhalo}
        \end{table}

        \begin{table}
            \hspace{2.0cm} log M$_{*} < 9.9$ \hspace{1.0cm} log M$_{*} >  9.9$  \\
          \begin{tabular}{c c c c c c c}
         
            \hline
            Redshift & $a$  & $b$ & $c$ & $a$ & $b$ & $c$ \\
            \hline
              0.4  & 0.41 & 0.07 & 6.89 & 1.23 & 0.09 & -1.31 \\
              0.7  & 0.41 & 0.09 & 6.89 & 1.29 & 0.11 & -1.86 \\
              1.0  & 0.41 & 0.11 & 6.90 & 1.35 & 0.13 & -2.40 \\
              1.3  & 0.41 & 0.14 & 6.94 & 1.33 & 0.16 & -2.32 \\
              1.6  & 0.40 & 0.17 & 6.98 & 1.36 & 0.20 & -2.59 \\
              1.9  & 0.39 & 0.21 & 7.03 & 1.38 & 0.24 & -2.91 \\
            \hline
          \end{tabular}
        \caption{The fitted values of a, b and c for the relation between stellar mass, halo mass, and $t_{\rm form}$ as found through the Galacticus simulation. These values are used in equation (9).  }
        \label{tab:stellarhalo}
        \end{table}

        \begin{table}
            \hspace{2.0cm} log M$_{*} < 9.9$ \hspace{1.0cm} log M$_{*} >  9.9$  \\
          \begin{tabular}{c c c c c c c}
         
            \hline
            Redshift & $a$  & $b$ & $c$ & $a$ & $b$ & $c$ \\
            \hline
              0.4  & 0.32 & 0.28 & 5.39 & 0.11 & 1.14 & -0.97 \\
              0.7  & 0.36 & 0.25 & 5.31 & 0.13 & 1.22 & -2.02 \\
              1.0  & 0.37 & 0.24 & 5.33 & 0.20  & 1.23 & -2.86 \\
              1.3  & 0.38 & 0.23 & 5.38 & 0.33  & 1.13 & -3.09 \\
              1.6  & 0.38 & 0.23 & 5.38 & 0.41 & 1.10 & -3.51  \\
              1.9  & 0.38 & 0.23 & 5.36 & 0.50 & 1.00 & -3.44 \\
            \hline
          \end{tabular}
        \caption{The fitted values of a, b and c for the relation between stellar mass, halo mass, and dynamical mass, as found through the Galacticus simulation. These values are used in equation (10).  }
        \label{tab:stellarhalo}
        \end{table}

\subsection{Halo Masses from Dynamical Masses using Models}
        
\subsubsection{Methods and Results}

Our goal in this section is to examine the relation between the observed dynamical mass 
and halo masses of galaxies at $z < 3$.    We furthermore also investigate the use of the
S$_{0.5}$ parameter for measuring the halo masses of galaxies.

One reason for investigating this in detail is that based on the results 
of \S 4.2, the scatter in the stellar mass to halo mass relation is reduced by including the 
dynamical masses, ages and halo concentrations in the inference of halo mass from the stellar mass.
It might be the case, and indeed we later continue to show, that the relation between M$_{\rm dyn}$ and M$_{\rm halo}$
has a lower scatter and cleaner correlation than between halo mass and stellar mass.
  
Based on this we make the assumption that the dynamical mass is a better indicator of the 
halo mass than the stellar mass, as not only does it have a reduce scatter, but also it is a 
measure of the internal motions of galaxies, and thus more directly aligned with a 
measurement of the halo properties than the stellar mass, which is based on complicated 
baryonic physics and star formation histories (e.g., Hearin \& Watson 2013).

We investigate this using both the Galacticus and Millennium simulation results.  
We find that the two models give essentially the same pattern between the dynamical 
mass and halo mass.  We use
the values of the half-mass radius, velocity dispersion and maximum rotational 
velocity from the Galacticus simulation to determine the relation between 
M$_{\rm dyn}$ and M$_{\rm halo}$ using the calculation of M$_{\rm dyn}$ from 
eq. 2, as well as when we use the V$_{\rm max}$ value instead of S$_{0.5}$.  

When we compare the relationship between M$_{\rm dyn}$, as measured 
using S$_{0.5}$, and the halo mass we obtain Figure~5.   The best fitting 
relation between these parameters is given by:

\begin{equation}
          \log(\mbox{M}_{\rm{halo}}) = \gamma \, \log \left(\mbox{M}_{\rm{dyn}}\right) + \delta. \\
\end{equation}

\noindent Where we list the values for $\gamma$ and $\delta$ in Table~8.  As can be seen, there is very little evolution in redshift in these parameters, and in fact on average the predicted dynamical mass to total mass relation does not change much up to $z = 2$ in these models.     

We now ask the question of whether it is better to use these dynamical masses or stellar masses to obtain M$_{\rm halo}$.  Essentially, we want to derive M$_{\rm halo}$ from observations, and it is likely that M$_{*}$ or some form of dynamical mass is the best way to do this.    For a method of deriving M$_{\rm halo}$ to work well we want a maximum sensitivity, i.e, as you move along the `observed' axis, small changes produce a significant amount of change in the halo mass measure.   Flat relations like that seen for the sigma mass in \S 4.3.3 have 'shallow' fits where similar halo masses are retrieved for a range of `kinematic' masses.   However, we also do not want a very steep relationship, such that small changes in observables produce large changes in halo masses. This is because all observables have uncertainties, and inaccuracies are magnified by a very steep relationship.  

Ideally, we want something which has a 1:1 slope as much as possible,  and this is provided more by the relation between dynamical vs. halo masses than by the stellar vs. halo mass relations.   This can be seen by the slopes in the M$_{\rm dyn}$ vs. M$_{\rm halo}$ relations shown in Table~8 which have values of $\sim 1.2-1.3$.   This way, galaxy halos can be derived with maximum sensitivity, without inducing large errors. Note that this criteria for deriving maximum usability is in concert with minimizing the scatter, which M$_{\rm dyn}$ vs. M$_{\rm halo}$ also does (Figure~3).

 Therefore we conclude that
using a dynamical mass is superior to using a stellar mass to obtain the halo mass.  Using the Moster et al. (2010) formalism (see Appendix) we are also able to retrieve a good fit, particularly for the lower mass galaxies.   We hereafter use this  as the model method for finding the total halo masses of galaxies based on their dynamical masses as measured from the kinematics and sizes of 
galaxies.

        \begin{table}
          \begin{tabular}{c c c}
           
            \hline
            Redshift & $\gamma$  & $\delta$   \\
            \hline
              0.4  & 1.34$\pm$0.02 & -1.80$\pm$0.26 \\
	      0.7  & 1.21$\pm$0.02 & -0.43$\pm$0.23 \\
              1.0  & 1.29$\pm$0.04 & -1.22$\pm$0.40 \\
	      1.3  & 1.21$\pm$0.02 & -0.43$\pm$0.23 \\
              1.6  & 1.20$\pm$0.03 & -0.30$\pm$0.26   \\
	      1.9  & 1.21$\pm$0.05 & -0.45$\pm$0.45 \\
            \hline
          \end{tabular}
        \caption{The fitted values of $\alpha$ and $\beta$ for the relation between dynamical mass and halo mass using the results of the Galacticus simulation.   These values are used in Equation (5), and are generally best used for systems with larger stellar masses, e.g., log M$_{*} > 10$.  }
        \label{tab:stellarhalo}
        \end{table}

However, it must be noted that in a real sense these `measures'  are model dependent, and as such when we compare with other methods of measuring
the halo mass we are in a sense testing this methodology for measuring the halo masses of
galaxies based on observational features.    This methodology is thus not ideal, as it relies on semi-analytical models, and is not based on fundamental observations or derivations.   However, if the mass profile of real galaxies is similar to those in the models we use, our methodology should be effective at tracing the halo masses of individual galaxies.  Importantly, this method is superior to just using
the stellar mass.   

This is also not fundamentally different from halo masses measured through abundance matching, or clustering, where there is an assumption about dark matter halo masses and how these are distributed in abundances and clustering.   Alternative methods of deriving and their comparisons between the total masses from lensing or the kinematics of large radii tracers such as clustering will be addressed below.

\subsubsection{Comparison to Virial Mass at Virial Radius }

Another direct way to measure the total or halo mass of a galaxy is to use, for disk like galaxies,
the circular velocities and virial radius to make a measurement of the halo mass.  This can be done using theoretical
arguments which relate observed quantities of the effective radius to the virial radius, as well as the 
circular velocity to the V$_{\rm max}$ values which are measured directly from the data.  

This can be done in the following way, as first outlined in a similar way in Lampichler et al. (2017).  
The relationship between the effective radius and the virial radius is given by Kravtsov et al. (2013) and Agertz \& Kravtsov et al. (2016)
using using the abundance matching assumption that $n_{\rm h} (>M) = n_{g}(>M_{*})$ such that:

$$R_{200} = 66.67 r_{1/2}.$$

\noindent This equation however requires us to know the half-mass radius, as opposed to the half-light
or effective radius, which is what we have measured for the bulk of our galaxies.  However, when 
the half-light radius is compared to the half-mass radius, using the mass maps of galaxies, there is little
systematic difference found.  Therefore we use the value of the effective radius when measuring the virial
mass through this method, i.e., $r_{1/2} = r_{\rm eff}$ 
(e.g., Lanyon-Foster et al. 2012).

The next step when converting observations to virial masses through this method requires that we use the
circular velocity, rather than the maximum velocity.   For this we find that the circular velocity to 
maximum velocity ratio is given by Cattaneo et al. (2014) such that 
$v_{\rm circ} = 1.33 \times V_{\rm max}$.  We compare our masses measured through this approach using the equation:

\begin{equation}
{\rm M_{circ} = \frac{V_{\rm circ}^{2} \times R_{1/2}}{\rm G} = \frac{1.33 \times V_{\rm max}^{2} \times R_{1/2}}{\rm G}},
\end{equation}

\noindent which
we call the circular masses.  We compare these masses with those obtained through our direct approach.  The result of this 
is shown at three different redshifts in Figure~6, where there is a good overlap with the mass measurements using our approach.  This is particular the case for those systems which are rotationally dominated with $v/\sigma > 1$, as shown by the blue points in Figure~6.  The analytical relation goes through these points expect at the highest masses.

\subsubsection{Sigma Masses - Early Type Galaxies}

A popular method for obtaining the total masses of early type galaxies or ellipticals that are
dominated by their velocity dispersion is to use the formula:

\begin{equation}
{\rm M_{\rm sig} = \frac{5 \times \sigma^{2} \times R_{\rm e}}{\rm G}},
\end{equation}

\noindent which we call the sigma ($\sigma$) mass.   This is often
used to determine the total masses of galaxies, even at high redshifts (e.g., Treu
et al. 2005).  We show in Figure~7 the
relationship between the sigma mass and the halo mass found 
within the Galacticus models.  We find that the sigma mass is a good tracer of halo
mass at the highest masses, where log M$_{\rm sig} > 12.5$, but note that the highest mass end
of this relation is dominated by galaxies with large sizes.

We also plot this ratio between the halo
mass and the sigma mass on the right hand panel of Figure~7.    This demonstrates that the
halo mass of the highest mass elliptical galaxies can be retrieved from the measurement of internal
velocity dispersions ($\sigma$) for most, but not all of these high-$\sigma$ galaxies.

We furthermore find that the relation between the halo mass and the sigma mass is not well fit by
a power-law between the two, as we find for the halo mass and the dynamical S$_{0.5}$ mass.  In fact, we
find a good fit when we use the formalism between halo and stellar mass found by Moster et al. (2010) derived from abundance match samples.  We show this best fit as the red line in Figure~7.  We discuss this
relation in more detail in the Appendix.

\subsection{Abundance Matching Masses}

We further investigate the halo mass to stellar mass ratios from halo abundance matching using
stellar masses calculated using the same methodology and underlying techniques as we do for
the primary sample's stellar masses.  The basic idea behind this method is to use measured stellar mass number densities at various redshifts and dark matter relation predictions to determine the halo masses for systems with the same abundances  (e.g., Kravtsov et al. 2004; Shankar et al. 2006; Conroy et al. 2007; Moster et al. 2013; Behroozi et al. 2013; Shankar et al. 2014; Buchan \& Shankar 2016).  

 Our primary method is to use the number densities for galaxies as a function of redshift as derived by Mortlock et al. (2015), who furthermore use the Chabrier IMF and fitting methods that we use for our measured stellar masses when comparing with kinematic dynamical masses.  We then match these number densities, which get higher at lower masses, to that of dark matter halo abundances at the same redshifts and thereby associate each mass range with a halo mass range.
        
To carry out this comparison, the mass function of dark matter halos (including sub-halos) is assumed to be monotonically related to the observed stellar mass function of galaxies with zero scatter.  This relation is given by,

\begin{equation}
n_g(>M_{\rm star})=n_h(>M_{\rm halo})
\end{equation}

\noindent where, $n_g$ and $n_h$ are the number densities of galaxies and dark matter halos, respectively. 

To derive these we use the Jenkins et al. (2001) modification to the Sheth \& Tormen (1999) halo mass function, the analytic halo model of Seljak (2000), and generate the linear power spectrum using the fitting formulae of Eisenstein \& Hu (1998).  Based on this the predicted number density of dark matter halos is given by,

\begin{equation}
n_h(>M_{\rm halo})=\bar{\rho}\int_{M_{min}}^{\inf}\frac{<N>}{M_{\rm halo}} f(\nu)\, d\nu.
\end{equation}

\noindent Where $f(\nu)$ is the scale independent halo mass function, $\nu=[\delta_c/\sigma(M_{halo})]^2$ ($\delta_c=1.68$ is the value for spherical over-density collapse). $\sigma(M_{halo})^2$ is the variance in spheres of matter in the linear power spectrum, $\bar{\rho}$ is the mean density of the Universe, and $<N>$ is the average number of halos, including sub-halos where we assume the fraction of sub-halos (f$_{sub}$) is described by,

\begin{equation}
f_{sub}=0.2-\frac{0.1}{3}z,
\end{equation}

\noindent as in Conroy \& Wechsler (2009).  Note that this fraction is small and therefore has little effect on our results, but we include it for completeness. Our method basically assigns the most massive galaxies, as measured in stellar mass, to the most massive halos.  Given its simplicity it is reasonably successful at matching various observations at multiple epochs (e.g., mass-to-light ratios, clustering measurements, see Conroy \& Wechsler 2009 and references therein).

We find that the fits between the stellar mass and halo mass with abundance matching is well represented by the following analytical functions at $z = 0.40, 0.75, 1.25, 1.75, 2.25$ and $2.75$ as:

        \begin{table}
          \begin{tabular}{c c c c}
         
            \hline
            Redshift & $a (\times 10^{-10})$  & $b$ & $c$  \\
            \hline
              0.40  & 3.68$\pm$0.25 & 9.27$\pm$0.28 & 11.22  \\
              0.70  & 2.68$\pm$0.29 & 9.41$\pm$0.28 & 11.22  \\
              1.25  & 2.73$\pm$0.38 & 8.53$\pm$0.16 & 11.22   \\
              1.75  & 8.28$\pm$0.35 & 8.00$\pm$0.05 & 11.22  \\
              2.25  & 2.69$\pm$0.34 & 8.46$\pm$0.05 & 11.22   \\
              2.75  & 3.64$\pm$0.77 & 6.39$\pm$0.08 & 11.22  \\
            \hline
          \end{tabular}
        \caption{The fitted values for eq. 17 which relates the stellar mass and halo
mass for galaxies using abundance matching (\S 4.3). }
        \label{tab:stellarhalo}
        \end{table}

\begin{equation}
{\rm log\, M_{halo}} = a ({\rm log\, M_{*}})^{b} + c
\end{equation}

\noindent which are all fits with low $\chi^{2}$ values. The results of these fits are shown in 
Table~9. This differs slightly from other parameterizations
which include up to five parameters (Behroozi et al. 2013), yet this is the
simplest form that fits our data.

We compare the relation between the halo masses derived from abundance 
matching  and the stellar masses at the same limits in Figure~8. The
halo masses derived from abundance matching are derived based on the most massive galaxies 
at each 
corresponding stellar mass. Thus it is possible, and likely even, that galaxies which are within
larger halos are found to higher a higher halo mass at lower stellar mass through abundance 
matching than what we find through the model masses derived in \S 4.1.

We show 
this comparison at redshifts from z = 0.4 to z = 3.  First, we note that 
there is a small amount of evolution in the halo to stellar mass ratio 
as a function of redshifts, which is also what we 
find when we investigate this relationship using the observable relations from 
simulation output (\S 4.1).    Although we note that at $z > 1.5$ the 
relation is higher, such that the halo mass is larger at a given stellar mass, 
particularly at lower masses.  At higher masses the trend is not as clear.

We also overplot on Figure~8 the relation between halo and stellar mass derived
from the model masses derived from stellar mass (eq. 5, \S 4.1) at redshifts
$z \sim 0.4$ and $z \sim 1.0$.  There are some differences between the 
abundance matched based masses and the stellar mass based model masses, 
particularly at the lower mass range, however this is 
due to the fact that there is unlikely a 1:1 galaxy:halo ratio at these
masses, something we investigate and discuss further below.
Hence we get an overestimate of the individual halo masses for these
systems using abundance matching. This is not an unexpected result to 
some degree within this formalism, and this
effect has been seen before by e.g., Conroy \& Wechsler (2009).  
We compare the calculated halo masses as a function of stellar mass using 
the model approach from eq. (5) to those from abundance matching in 
Figure~9, which shows the ratio of the halo model masses vs. 
the abundance match masses.

Overall, we find that the average difference between the two methods of measuring M$_{\rm halo}$ 
is less than a factor of three at log M$_{*} >$ 10.5, which is the primary mass range we study here. 
This difference 
is such that the abundance matching masses overestimate halo masses by a factor of roughly 1.5-3 
compared to dynamically based halo masses at log M$_{*} < 11$.  The halo masses are similar within these two methods at the highest masses.  However, at log M$_{*} < 10$ the difference is large, a 
factor of $\sim 4$, between the abundance match based masses and those from the models.  This is likely due to abundance matching
locating these lower mass galaxies in overall higher halos as few systems are isolated.

This shows that our method of measuring the halo masses for massive galaxies through models is roughly consistent with other independent 
methods at the highest masses, and thus likely to be reliable given that  halo abundance matching 
reveals nearly the same halo masses for a given stellar mass in the range we are 
interested, with an uncertainty similar to that of measuring the stellar masses.  We furthermore
compare our results to other abundance matching results, as well as with gravitational 
lensing ones in \S 5.

\section{Observational Results}

In this observational results section we first describe the relations between the various masses 
calculated in \S 3 and \S 4 for actual galaxies, including using our new derived individual halo masses.  
This is followed by a discussion of how these results can be applied towards understanding how 
galaxy assembly occurs in terms of its major mass components.  Later,
in the discussion section we discuss the implications of these results.

  \subsection{Mass Scaling Relations}

    In this section we use the various measured and calculated 
stellar, dynamic, and halo masses to link the stellar and dark masses of galaxies up 
to $z \sim 1.2$ for our mostly star forming primary sample. This includes 
investigating how stellar mass relates to these other masses observationally.   Ultimately, 
we are interested in applying these relations to data to infer galaxy evolution, which we discuss in 
\S 6..

        \subsubsection{The Tully-Fisher Relation}
          
        \begin{figure*}
          \centering
\vspace{-2cm}

          \includegraphics[angle=-90, width=19cm]{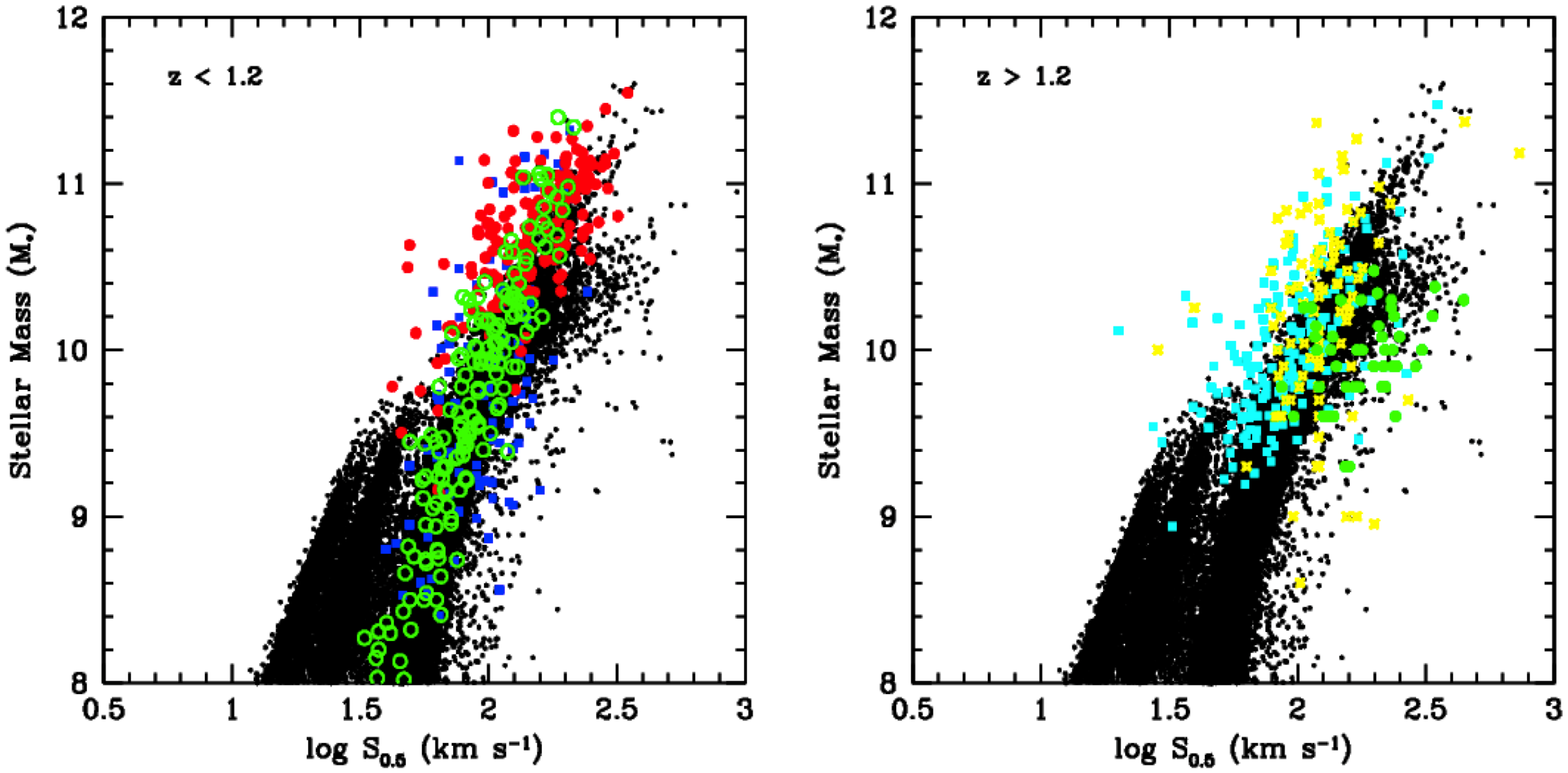}
	\vspace{-2.5cm}
          \caption{The S$_{0.5}$ stellar mass Tully-Fisher relation for our sample of galaxies, as well as for the Galacticus models which we use to calculate the halo masses from dynamical masses.  Shown on the left are the values for the $z < 1.2$ galaxies, whilst the right hand side shows the values for galaxies at $z > 1.2$.   On the left hand side, the blue small boxes are for the disk galaxies from Conselice et al. (2005), the green circles are from the disk galaxies in Miller et al. (2011, 2014),  the red circles are for the ellipticals at $z < 1.2$ taken from Treu et al. (2005).  For the higher redshifts on the right, the cyan squares are for the MOSDEF data from Price et al. (2016), the yellow crosses are for Erb et al. (2008), and the solid green circles are from Forester-Schreiber et al. (2006).  }
          \label{fig:S05MbaryTFR}
        \end{figure*}

For a first step towards investigating evolution of the different dark masses in galaxies, we construct a Tully-Fisher relation (hereafter TFR) for our sample and compare this to the simulation data that we use to convert dynamical into halo masses. Using the primary sample in this paper, Kassin et al. (2007) use the \SK\ parameter to construct a new type of  stellar mass Tully-Fisher relation which takes into account the ionized gas velocity dispersion (\SK/\Mst TFR; hereafter kinematic TFR).    The form of this Tully-Fisher relation given by:

$$M_{*} = a \times {\rm log S_{0.5}} + b$$

\noindent where $a$ is the slope of the relation and $b$ is the intercept.  This kinematic TFR is considerably tighter than TFRs calculated without taking into account the velocity dispersion, and has no detectable evolution over the redshift range $0.1 < z < 1$.  

We show the S$_{0.5}$ Tully-Fisher relation in Figure~10 at redshifts $z < 1$ and at $z > 1$.  As can be seen there is a good relationship between the stellar mass and the values of S$_{0.5}$ at both high and low redshifts. This involves galaxies of different types.  Furthermore, we also plot on these relations the prediction between these two quantities at both high and low redshift from the Galacticus simulation output.  As furthermore can be seen, we find a good agreement with the location of our data and theoretical model output, demonstrating that we can use these simulations as a good cosmological representation of all galaxies up to $z \sim 3$.

        \subsubsection{Dynamical and Stellar Masses}
          \label{sec:resultsDyn}
   
        \begin{figure*}
          \centering
\vspace{-2cm}

          \includegraphics[angle=90, width=19cm]{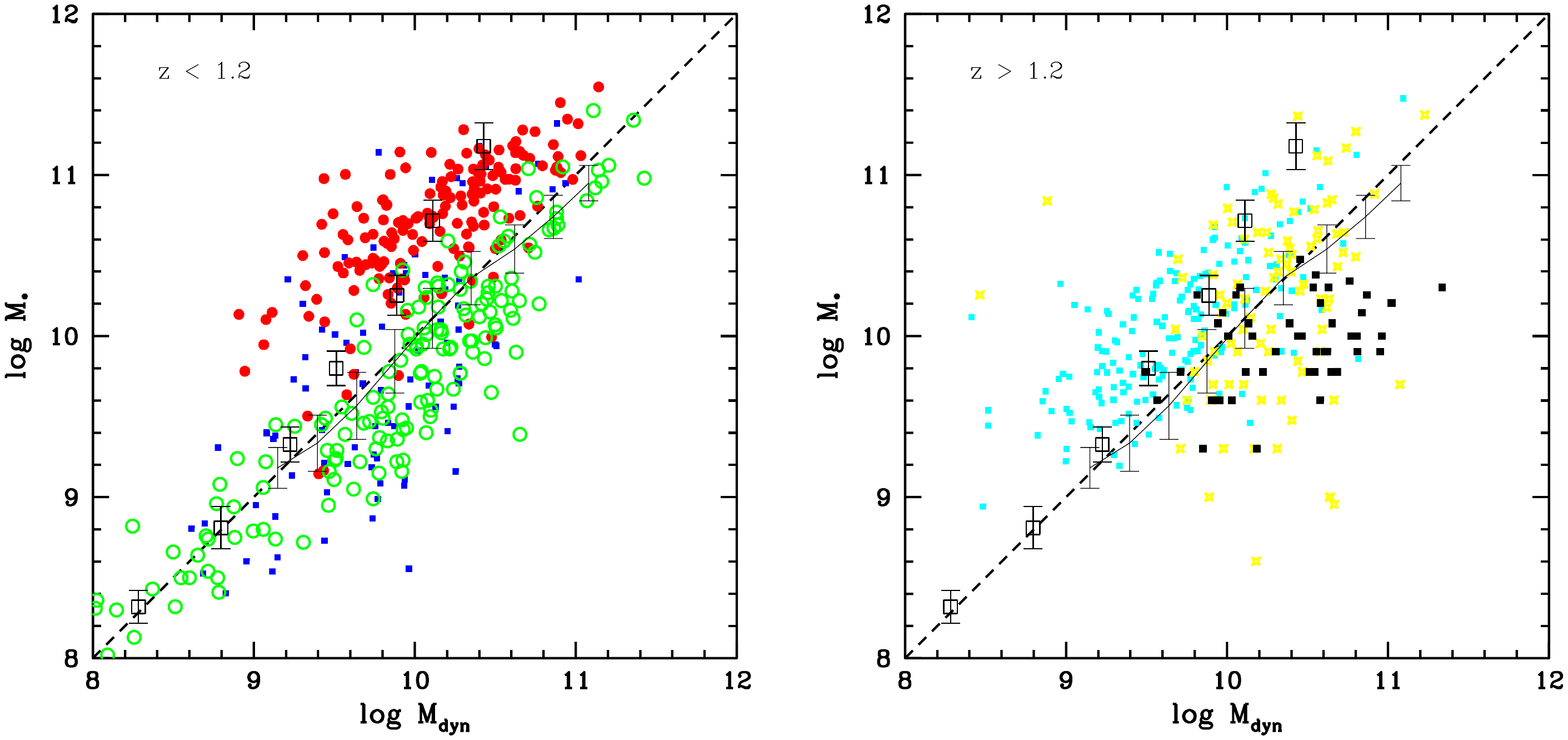}
	\vspace{-2.5cm}
           \caption{Figure showing the relationship between the
 stellar and dynamical mass divided into two redshift bins with the left hand
panel for galaxies at $z < 1.2$ and the right hand panel shows $z > 1.2$ systems. The open boxes with error bars show the average and scatter of the primary sample.  The blue small boxes are for the disk galaxies from Conselice et al. (2005), the green circles are from the disk galaxies in Miller et al. (2011, 2014),  the red circles are for the ellipticals at $z < 1.2$ taken from Treu et al. (2005).  For the higher redshifts on the right, the cyan squares are for the MOSDEF data from Price et al. (2016), the yellow crosses are for Erb et al. (2008), and the 
solid boxes are from Forester-Schreiber et al. (2006).   A key to this is shown
in Figure~13.
The solid line with error-bars shows the model relationship between these two
quantities as derived in models (see Appendix A).   }
\end{figure*}

        \begin{figure}
          \centering
\vspace{-2cm}

          \includegraphics[angle=0, width=8.5cm]{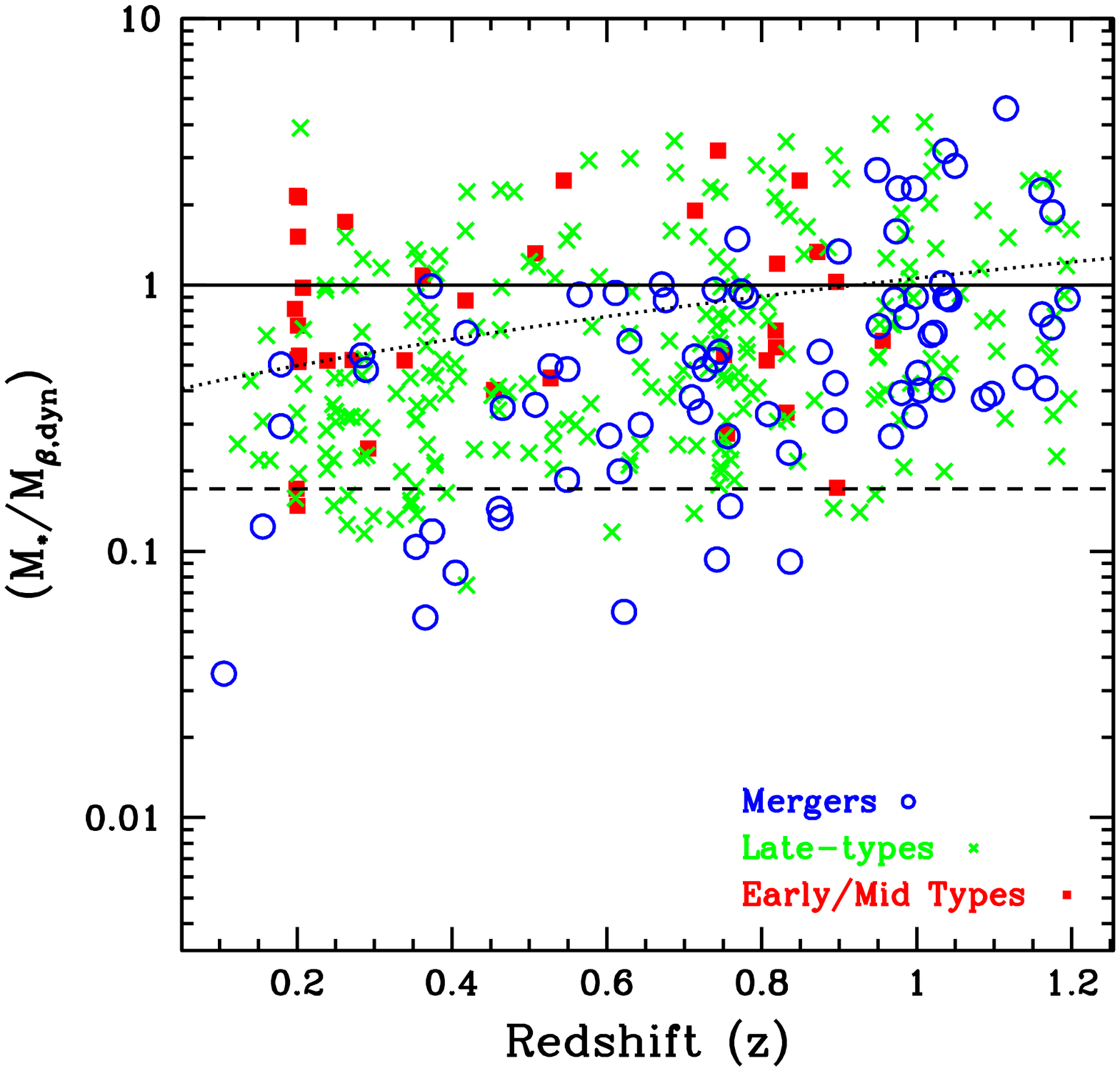}
\vspace{-2cm}
          \caption{The stellar to $\beta$, dynamical mass (eq. 3) ratio with respect to redshift for the primary sample.  The points are colored by the CAS morphological type.  There is some evidence for a change in the ratio with redshift, although we do see a stronger trend in the differences between morphological types, such that mergers and disks have a lower stellar mass to dynamical mass ratio than early-type galaxies.  The straight dot-dash line is the universal baryonic ratio and the dotted line is the best fit to the data.}
          \label{fig:Conselice05a}
        \end{figure}
          
        \begin{figure*}
          \centering
	\vspace{-2cm}
          \includegraphics[angle=-90, width=20cm]{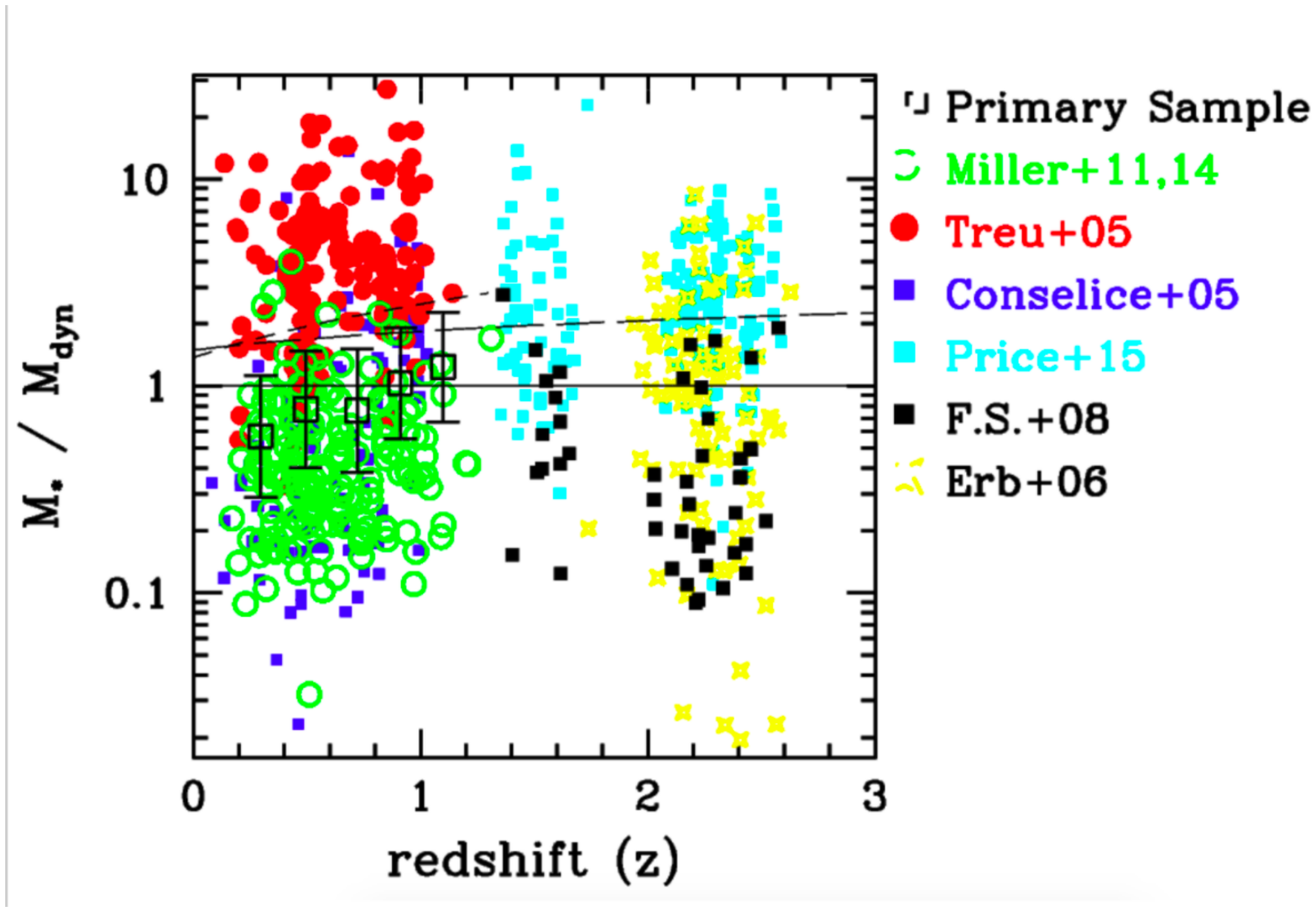}
\vspace{-2cm}  
        \caption{The relationship between the ratio of dynamical to stellar mass as a function
of redshift.  The colour of the points are the same as in Figure~11.  The long-dashed line shows the best fit to this relationship from low redshift to the highest redshifts at $z \sim 3$.   The short-dashed line shows the relationship between this mass ratio and redshift fitted up to $z \sim 1.3$.   We find little significant evolution at $z > 1$ in this relationship with the relation through all redshifts going as $\sim (1+z)^{0.3 \pm 0.12}$ (see text).}
          \label{fig:Conselice05a}
        \end{figure*}
          
In this section we compare the dynamical masses in our primary sample
from eq. (2) to their corresponding stellar mass measurements.  We plot the dynamical masses 
for the primary sample as a function of stellar mass in 
Figure~11.  We fit the stellar-to-dynamical mass 
relation for this primary sample as:
      
        \begin{equation}
            \log \left(\mbox{M}_{\rm{dyn}}\right) = (0.71 \pm 0.02)\, \log \left(\mbox{M}_{*}\right) + (3.17 \pm 0.17).
          \end{equation}

\noindent From Figure~11 it is 
clear that the dynamical mass, as we define it, within the visible radius is dominated by the 
baryonic and stellar mass.   As can also been seen in Figure~11, it is clear
that this ratio of dynamical to stellar mass does not change significantly
at different redshifts up to $z \sim 3$.

There is an interesting feature in the stellar to dynamical mass plot
which deserves a more careful look.  This is the fact that the stellar mass
for the primary sample, as well as the elliptical sample from Treu et al. (2005)
has the interesting property that the stellar mass is higher than the dynamical mass.  
This is certainly partially due to the way in which we define the dynamical mass and how it gives a better idea of the total mass for elongated and rotating systems than for compact systems such as ellipticals.  The reason for this
has to do with the fact that the total mass of spheroids is usually calculated with a constant in front of our dynamical mass equation, with most using `5' (see \S 4.3.3).  This is 0.7 dex, which is roughly the amount of the difference seen between the 1:1 ratio and the location of where ellipticals are located, both in the Treu et al. sample in our primary sample.

We furthermore use these relations 
to derive how the ratio of the stellar mass to 
dynamical mass changes with both dynamical and stellar mass.   
We find overall that these relations are indeed quantitatively
relatively constant as a function of redshift.  
Examining the relationship between
stellar and dynamical mass we find a best fit of:

\begin{equation}
{\rm log} \left(\frac{M_{*}}{M_{\rm dyn}}\right) = 0.41 \times {\rm log} M_{\rm dyn} - 4.64,
\end{equation}

\noindent and for the relation in terms of stellar mass we find,

\begin{equation}
{\rm log} \left(\frac{M_{*}}{M_{\rm dyn}}\right) = 0.29 \times {\rm log} M_{\rm *} - 3.17.
\end{equation}

\noindent Overall this reveals that the stellar mass ratio goes as 
$M_{*}/M_{\rm dyn} \sim M_{\rm dyn}^{0.41}$ and 
$M_{*}/M_{\rm dyn} \sim M_{\rm *}^{0.29}$.  This shows that within the 
visible extent of galaxies the efficiency of galaxy assembly, defined as having
higher M$_{*}$ to total mass, is larger at 
higher dynamical and stellar masses up to our limit of log $M_{*} \sim 11$ (Figure~1).  


We plot the ratio of stellar mass to dynamical beta mass against redshift in 
Figure~12 for our 
primary sample, divided into morphological types, and in Figure~13 for the total sample. 
We find that the evolution of this ratio with redshift over the entire sample can be fit as:

\begin{equation}
\left(\frac{M_{*}}{M_{\rm dyn}} \right) = (1.49\pm0.17)(1+z)^{0.30\pm0.12}.
\end{equation}

\noindent However, the bulk of this evolution occurs at $z < 1$.  If we carry out this fit
at $1 < z < 3$, we find a flat slope with (M$_{*}/$M$_{\rm dyn}) \sim (1+z)^{0.18\pm0.41}$.  On the other
hand we find an increase from $z \sim 0$ to $z \sim 1$ such that (M$_{*}/$M$_{\rm dyn}) 
= 1.37\pm0.60 \times (1+z)^{0.86+0.47}$.

This essentially means that there is very little to no evolution in the ratio of stellar to
dynamical masses at the highest redshifts.  A caveat must be put onto this however as our sample at 
higher redshifts are probably not representative.   However we are likely sampling the more gas rich
galaxies at these redshifts, and thus we would expect that value to increase, yet it remains statically
flat, at least in our overall sample.

        \begin{figure}
          \centering
\vspace{-2cm}

          \includegraphics[angle=0, width=8.5cm]{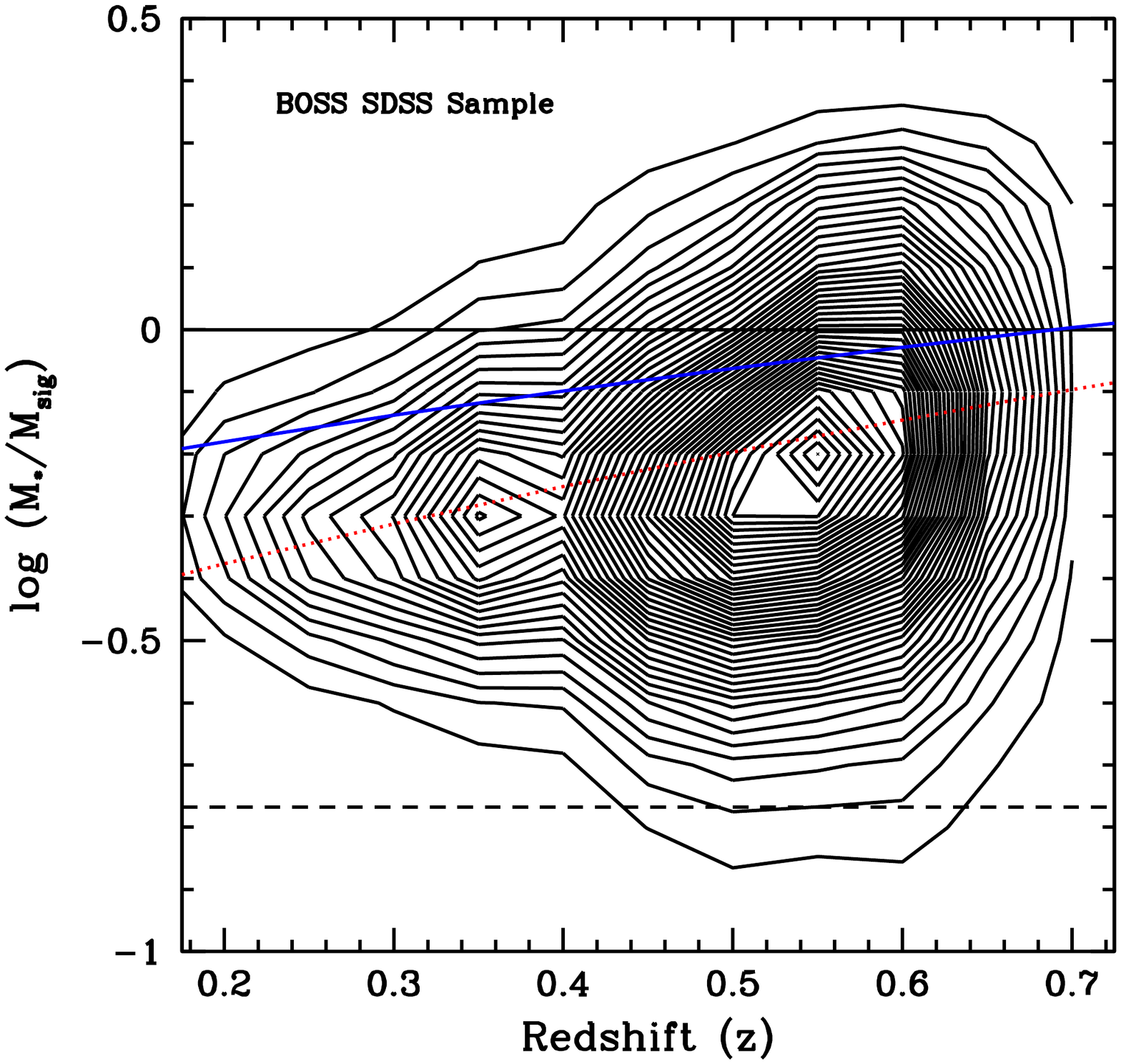}
\vspace{-2cm}

          \caption{Figure showing the relationship between the sigma masses and stellar masses
for the 180,000 galaxies in the BOSS sample as part of the SDSS sample in DR7 (Beifiori et 
al. 2014).  The red dashed line shows the best fit for the data, whilst the blue solid lines
show the best fit for the relation for our total sample but using the dynamical mass instead of
the sigma mass.}
          \label{fig:Conselice05a}
        \end{figure}

As a further way to test this, we investigate the BOSS sample from SDSS III, using galaxies
with high masses at M$_{*} > 10^{11}$ \solm, typically higher than those considered in this paper.   
In Figure~14 we show this relationship for galaxies from the BOSS sample
at $z < 0.7$. When
we do a least squares fit to this relationship between stellar to sigma mass ratio for
galaxies between M$_{*} = 10^{11} - 10^{11.5}$ \solm\, and redshift we find the relationship:

\begin{equation}
\left(\frac{M_{*}}{M_{\rm dyn}}\right) = (0.30\pm0.01)(1+z)^{1.85\pm0.1}
\end{equation}

\noindent where there is a slight increase with redshift in this ratio for the entire BOSS
sample.  This implies as well that there is some evidence for evolution in this ratio at $z < 1$, within the highest mass galaxies.
We later investigate this relationship for higher redshift systems 
in \S 5.2 using the  total halo masses of our systems.

Furthermore Figure~12 is also plotted by morphological type as measured 
by the CAS parameters (\S 2.2).  Based on this, we find that the early-type systems have a 
slightly higher stellar-to-dynamical mass ratio than other morphological types at all 
redshifts.    This is potentially a sign that the efficiency of galaxy formation is higher in early-type galaxies
than disks/late-types/mergers.   This is not a trend which results from the early type galaxies being more massive, as in our primary sample the ellipticals are not more massive on average than the other morphological types (Figure~1). 
This shows observationally that there is a third parameter 
in the relation of dynamical mass to halo mass, consistent with age as a third parameter
as discussed in \S 4.2.  This must be related to the galaxy 
morphology and time of formation which correlates with the concentration of the halo/galaxy.
We investigate this in more detail in \S 6.

      \subsubsection{Halo Mass  vs. Stellar Masses}

We now use these results to examine the relation between the halo and the stellar mass for our sample of real galaxies.   We show in Figure~15 the halo masses derived from \S 4 as a function of stellar mass for our entire observational sample.  The dashed black curved line shows the best three component theory fit from eq. (5), showing a reasonably good agreement with the data at the highest masses.    We attempt to improve the above fit and minimize scatter by investigating the stellar-to-halo mass fit when galaxies are binned by morphology and stellar mass density.   There is a slight improvement in these fits when the sample is binned by morphological type or by stellar mass surface density.

          \begin{figure*}
          \centering
\vspace{-2cm}

          \includegraphics[angle=90, width=17cm]{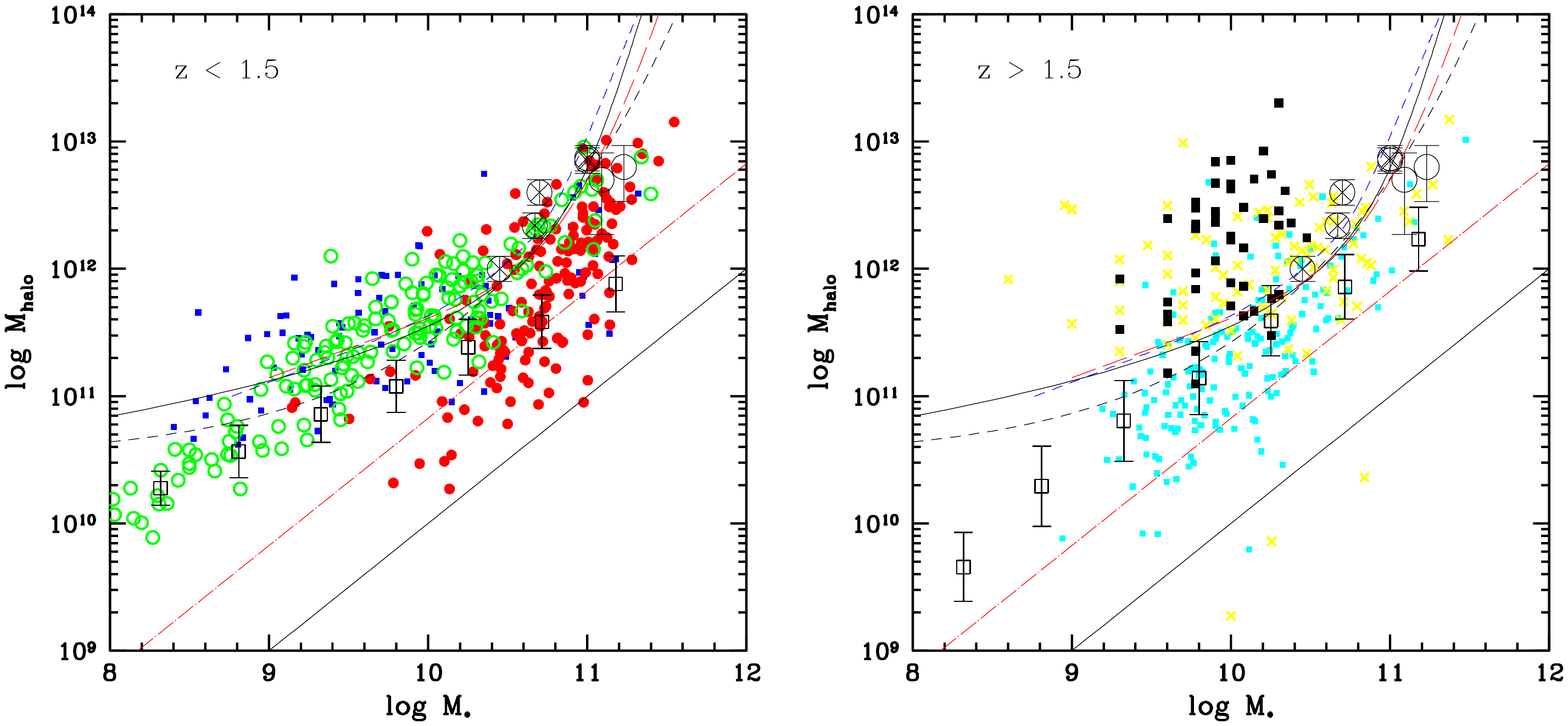}
\vspace{-2cm}

          \caption{Halo mass as a function of stellar mass computed using various samples and methods.   The open boxes with error
bars shows the average and dispersion for the primary sample of galaxies.  The large circles at log M$_{*} \sim 11$ are from the clustering analysis from Foucaud et al. (2010).  The points with circles and inner crosses are from the clustering measured masses of Skibba et al. (2015).  The black solid curved line is the relationship between stellar and halo mass from weak lensing measures from van Uitert et al (2016), the dashed red line is from the lensing results of Leauthaud et al. (2012), and the blue dashed line is from Moster et al. (2013).   The black dashed curved line, just below the dashed red line, is the relationship between stellar and halo mass as derived by the Galacticus simulation.  The solid straight line is the 1:1 relation between the two masses.  The other points are data with the same meaning as in Figure~11.  The dashed-dotted red line is the baryonic fraction from Planck. }
          \label{fig:MvirMstMbary}
        \end{figure*}

          \begin{figure*}
          \centering
\vspace{-2.5cm}

          \includegraphics[angle=90, width=17cm]{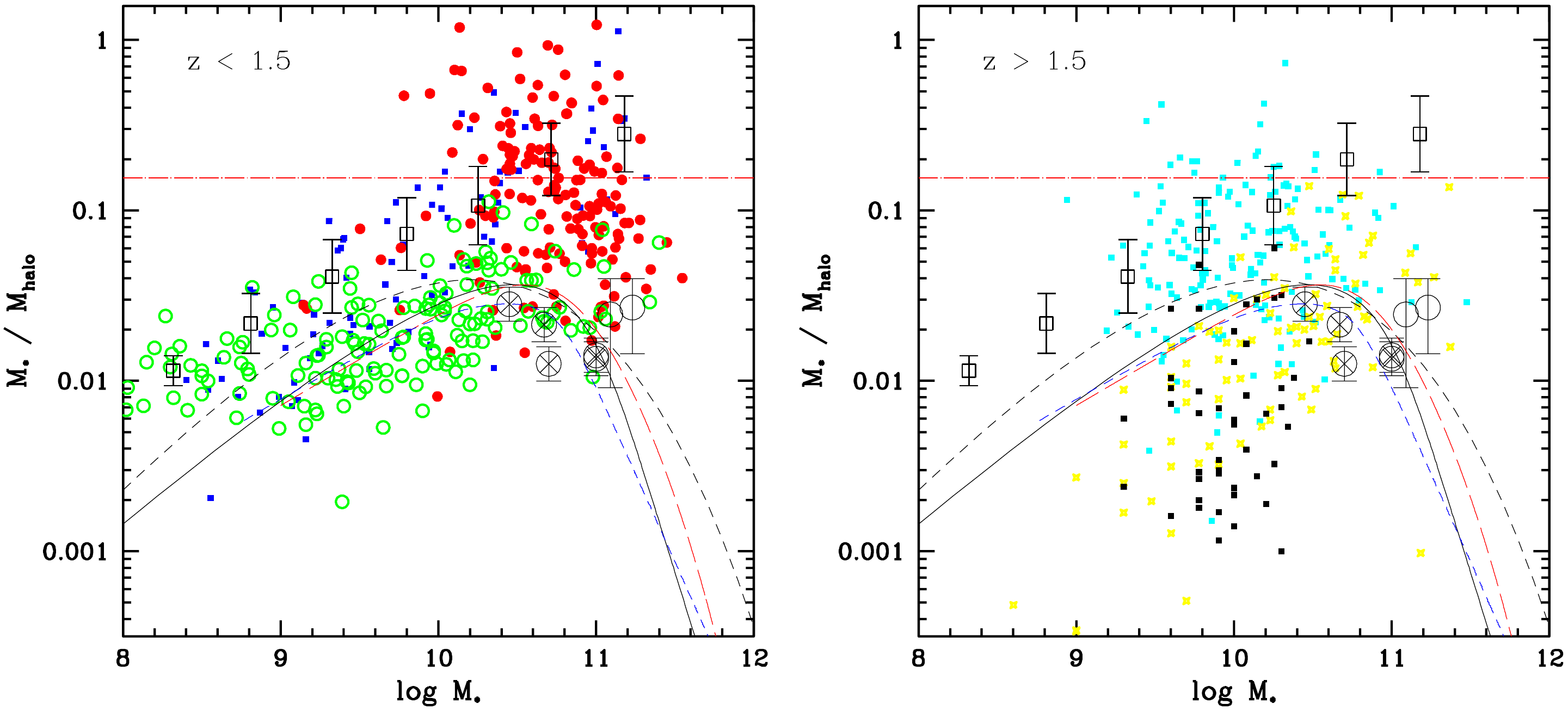}
\vspace{-2cm}

          \caption{The relationship between the stellar mass to halo mass ratio as a function of halo mass with a comparison to previous work.  The symbols and lines are the same as those explained in the caption to
Figure~15.  As can be seen we do not find a particular peak of galaxy formation efficiency, but a continual increase at higher stellar masses. }
          \label{fig:MvirMstMbary}
        \end{figure*}

        \begin{figure}
          \centering
\vspace{-2cm}

          \includegraphics[angle=0, width=8.5cm]{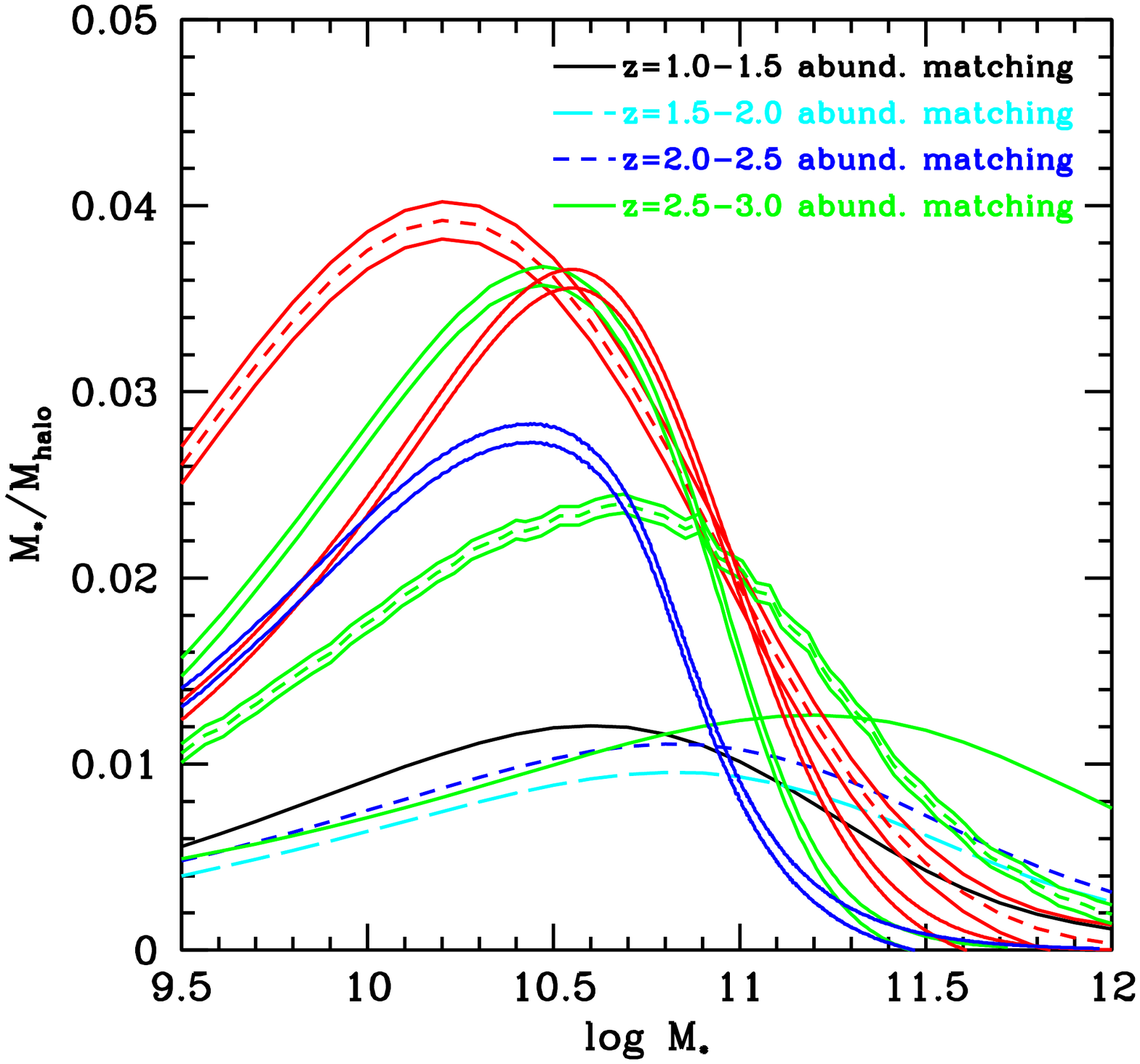}
\vspace{-2cm}
          \caption{The ratio of the stellar to halo mass derived through abundance matching and weak lensing.  Shown are four lines from the abundance matching method representing the relationship between these two quantities from $z = 1$ to $z = 3$.  The upper red dashed line surrounded by red solid lines shows the relationship between the halo and stellar masses as derived from the Galacticus simulation. The green double solid line is the relationship between stellar and halo mass from weak lensing measures from van Uitert et al (2016), the red double solid line is from the lensing results of Leauthaud et al. (2012), while the blue double line is from Moster et al. (2013).    We also show the abundance matching stellar and halo masses (green dashed line in the middle of two solid green lines) from Conroy and Wechsler (2009).}
          \label{fig:abundance}
        \end{figure}

Figure~16 shows the stellar mass to halo mass ratio as a 
function of stellar mass for our entire sample using both
primary and secondary sources.  To compare our halo 
masses with that of other work, we plot our results with 
those of Foucaud et al. (2010), 
van Uitert et al (2016), Leauthaud et al. (2012), and  
Moster et al. (2013).  These studies 
use different techniques to measure the halo masses of 
galaxies as described in \S 2.4.  The observational studies are shown as points of 
various shapes and colours.

        \begin{figure}
          \centering
\vspace{-2cm}

          \includegraphics[angle=0, width=7.4cm]{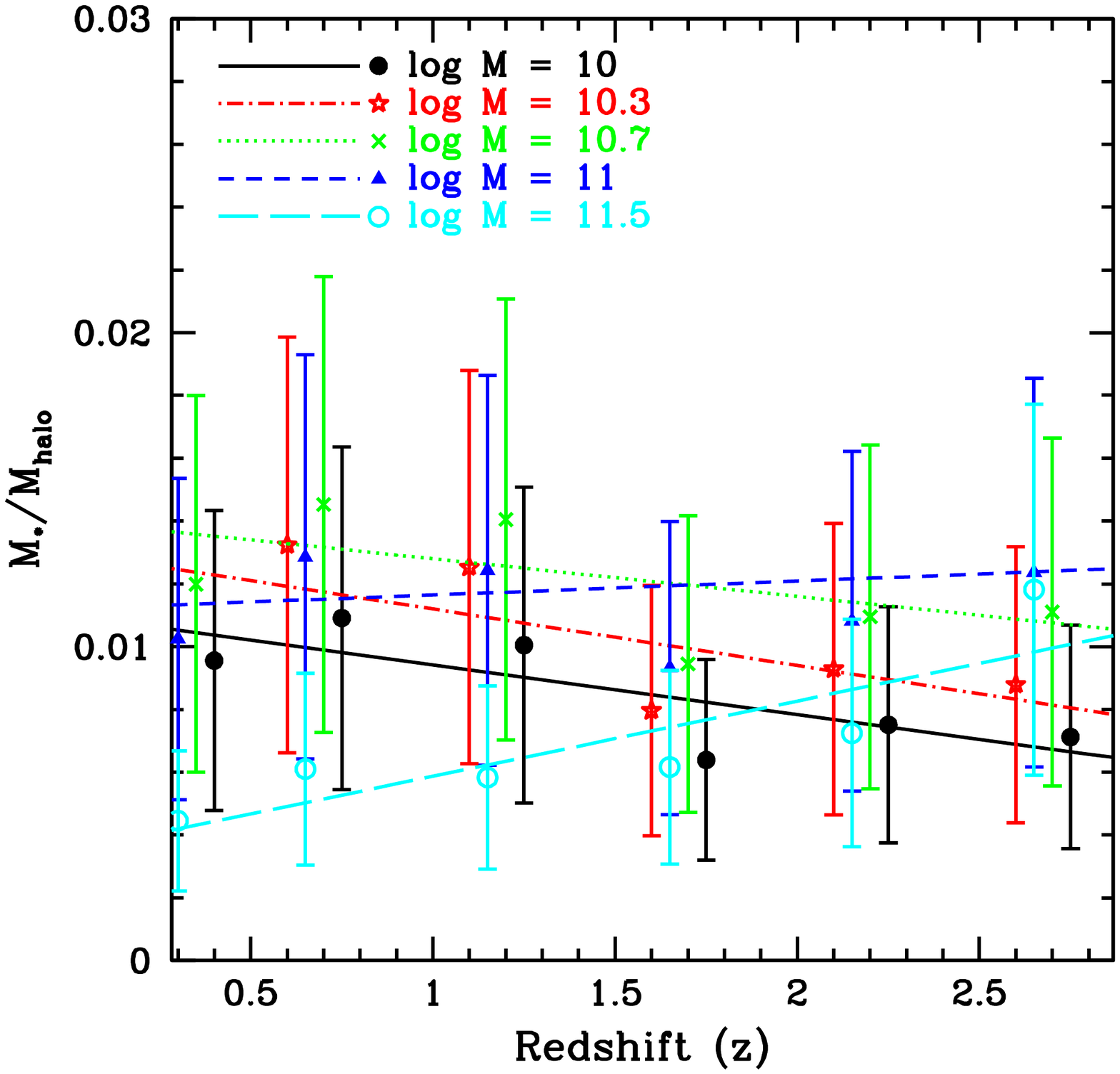}
\vspace{-2cm}
          \caption{The abundance matched derived stellar to halo mass ratios
as a function of redshift. Each line represents a selection in stellar mass.  As can be seen, there is very little evolution, and a formula fit suggestions that the slopes of these lines is not significantly different from a flat evolution (see text for details).}
          \label{fig:Sebz}
        \end{figure}

In the mass range $10^{10.5} <$ M$_{*}$/\msol\ $< 10^{12.0}$ we 
broadly agree with previous results, except for the Treu et al. (2005) 
ellipticals.  There is also an interesting feature in 
these figures such that there is no obvious turn-over in the 
\Mst/\Mtot\ ratio which reaches a maximum at about 
log M$_{\rm{halo}} = {11.5}$.   This turnover is seen 
in other samples  at this halo mass based on clustering
(e.g., Foucaud et al. 2010; Coupon et al. 2012).   For example,
the peak found by Coupon et al. (2012) it is log M$_{\rm halo}$ = 11.6
from clustering and in Coupon et al. (2015) is log M$_{\rm halo} = 12.2$. We
are able to probe down to log M$_{\rm halo} = 13$, yet we do not see an obvious
turnover in the ratio of stellar to halo mass to this limit.

    \begin{figure}
          \centering
\vspace{-2cm}

          \includegraphics[angle=0, width=8.5cm]{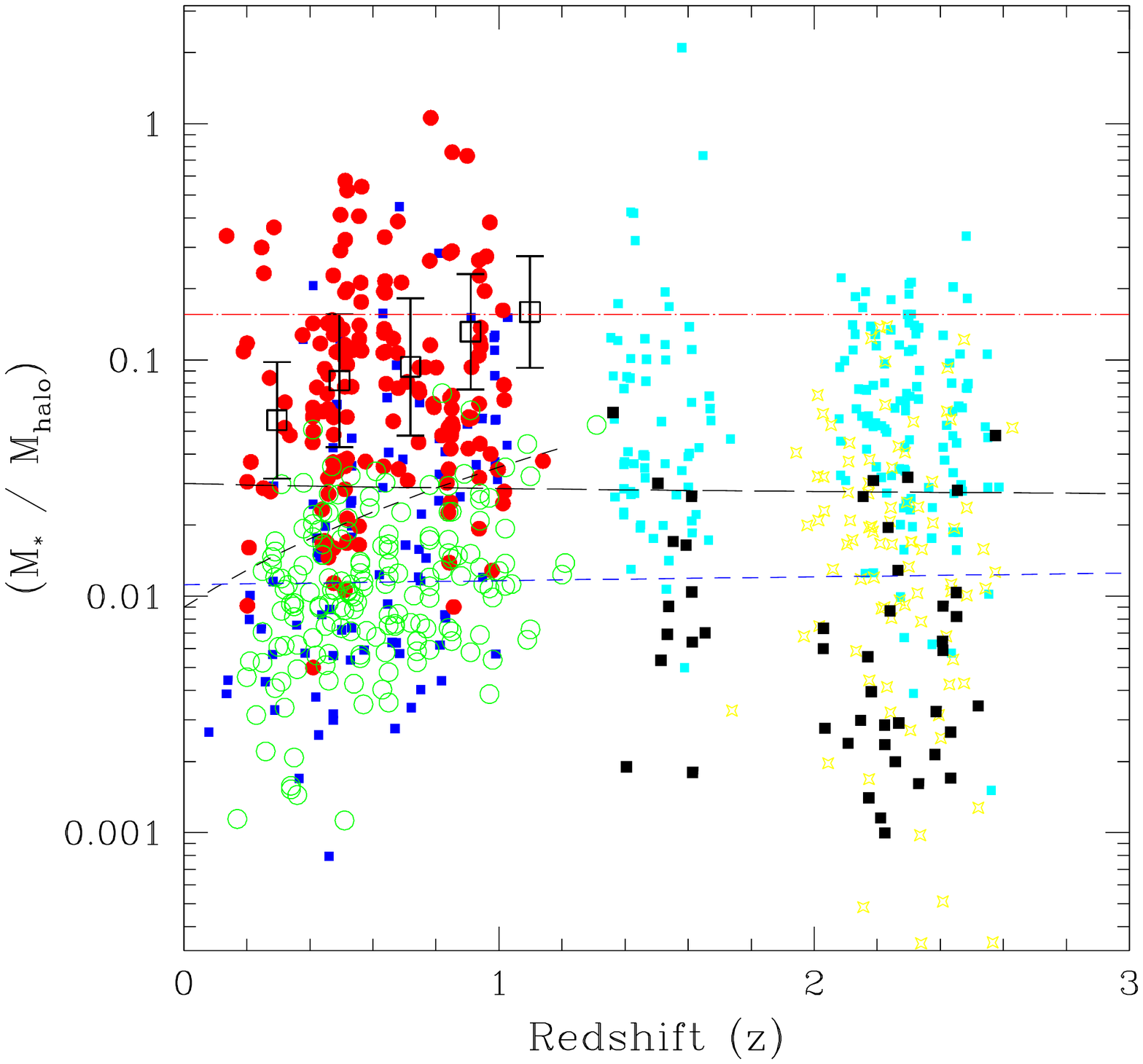}
\vspace{0.2cm}
\vspace{-2.5cm}

          \caption{The \Mst-\Mtot\ ratio ($f_{*}$) as a function of redshift.  The points and the red dash-dotted line are the same as Figure 11.  The red horizontal line shows the universal baryonic mass fraction from Planck.   We also show the best fit power laws of the
form $f \sim (1+z)^{m}$ (see text).  The long-dashed line shows this fit to all the data from z =0 to 3, while the short-dashed line shows the fit for sample at $z < 1.2$, where we find the exponent 
m = 1.97.  The blue dashed line shows the relationship between $f_{*}$ and redshift derived from the abundance matching (see text).  The points are otherwise the same as in
Figure~11. }
          \label{fig:Sebz}
        \end{figure}

This shows that 
the efficiency of galaxy formation is highest at the highest 
masses.    This relationship is 
also seen when ones compares the mass ratios by using halo 
masses as determined from abundance matching (e.g., Behroozi 
et al. 2013).  Figure~17 furthermore shows the ratio of our abundance
matching stellar vs. halo masses as a function of stellar mass.  
This demonstrates that the maximum galaxy formation efficiency is at
around log M$_{*} \sim 10.5$.   Finally, Figure~18 shows the evolution of the 
ratio of stellar to halo mass as a function of redshift, demonstrating little change 
with time, as discussed in the follow section.
   
    \subsection{Mass Ratios as a Function of Redshift}

One of the issues we investigate is how the ratio of stellar to halo mass varies with redshift.  This is however a difficult topic to study observationally and our attempts at answering this question should be seen as a preliminary solution until more kinematic and mass data is available for complete samples of galaxies at $z > 1.5$.  However, we can get some idea of this from limited observations of kinematics at various redshifts, as well as through our abundance matched masses.    

Overall, however, as we have kinematic data up to $z = 3$ we can determine the evolution of this relation up to these redshifts, although for a potentially biased sample of mostly star forming galaxies (\S 2.3).   Figures~12 \& 13 plot stellar-to-dynamical mass ratio as a function of redshift up to $z = 1.2$ for our primary and total sample, while Figure~19 plots the \Mst/\Mtot\ ratio against redshift for all of the observational samples compared with theory predictions and other observational results.    For
all samples of galaxies up to $z = 3$ we find a best fitting relationship of the form:

\begin{equation}
f_{\rm f,halo} = \left( \frac{M_{*}}{M_{\rm halo}}\right) = (0.07\pm0.01) \times (1+z)^{-0.07\pm0.11}
\end{equation}

\noindent This is consistent, within  1$\sigma$, of this ratio being  
flat, i.e., does not evolve with redshift.  We do find an evolution with 
a significant slope for the sample at $z < 1$ 
with $f_{\rm f,halo} = 0.028(1+z)^{1.97\pm0.29}$, as we do for the ratio of the
dynamical to stellar mass.    

This shows that the majority
of the evolution in terms of stellar vs. halo mass occurs at later times when
galaxies are primarily finished with their star formation, and are evolving more
in terms of galaxy mergers (e.g., Mundy et al. 2017).
In summary, we find no significant evidence  for any change in 
the \Mst/\Mdyn\ ratio at the highest redshift, although we do
find an evolving ratio at lower redshifts, $z < 1$.

We also find very little evolution in the stellar to halo mass ratio when 
using the abundance matching technique (\S 4.4).    We show in Figure~18
the evolution of the ratio of stellar to halo mass as derived from the 
abundance matching methodology.  Overall, we find that the 
stellar mass ratio does not significantly evolve with time at any measured 
stellar mass as measured either from the kinematic based model halo masses, 
or from using abundance matching.  

We fit the evolution of these mass ratios from abundance matching in the form of 

$$\left( \frac{M_{*}}{M_{\rm halo}} \right) = \alpha \times z + \beta.$$ 

\noindent The values of the 
best fitting parameters are $\alpha = -0.0016\pm0.0006$ and $\beta = 
0.011\pm0.001$ for galaxies with log $M = 10$, $\alpha = -0.0018\pm0.0008$, 
$\beta = 0.013\pm0.001$ for log $M = 10.3$ systems, $\alpha = -0.0012\pm0.0009$, 
$\beta = 0.014\pm0.001$ for log $M = 10.8$, $\alpha = 0.000045\pm0.000008$, 
$\beta = 0.011\pm0.001$ for log $M = 11$ and $\alpha = 0.0024\pm0.0007$, 
$\beta = 0.003\pm0.001$ for log $M = 11.5$.  With the only case which has a 
significant slope of $> 3\sigma$ is for the highest masses with log $M = 11.5$.

Our results are in agreement with Conselice 
et al. (2005), who also find that the highest mass galaxies have 
the highest \Mst/\Mtot\ ratio.   
Furthermore, using galaxy clustering, Foucaud et al. (2010) find a 
declining \Mst/\Mtot\ 
ratio with redshift at $z>1.0$, at a given stellar mass selection.  However, 
Foucaud et al. (2010) effectively measure the parent halo of objects 
while we are observing the halo mass for individual `sub-halos'.
Foucaud et al. (2010) interpreted this evolution as a halo downsizing effect.


However, we must keep in mind that the kinematic model mass evolution in the mass ratios is biased.  
The galaxies towards the higher redshifts are more typically star forming 
galaxies, with a later-type morphology, and there 
is no guarantee that these systems are representative of the population as a 
whole at these redshifts.   Our results at $z > 1.5$ should therefore be taken 
with caution as a preliminary measure of how 
total and stellar masses relate at higher redshifts.

  \section{Discussion}
    \label{sec:Discussion}

\subsection{Dark Matter Accretion Rates}

These results have implications for the way in which galaxy formation 
occurs at $z < 3$.  With the observation of a nearly constant stellar to 
total mass ratio with redshift we can derive the accretion rate of both gas, 
as well as dark matter,
building the halos and galaxies over cosmic time.  We start off with a method
for parameterizing the star formation history,  which for our sample can
be represented by an exponentially declining star formation, such that
the star formation rate as a given redshift $\Psi(z)$ can be written as

        \begin{figure}
          \centering

          \includegraphics[angle=-90, width=9cm]{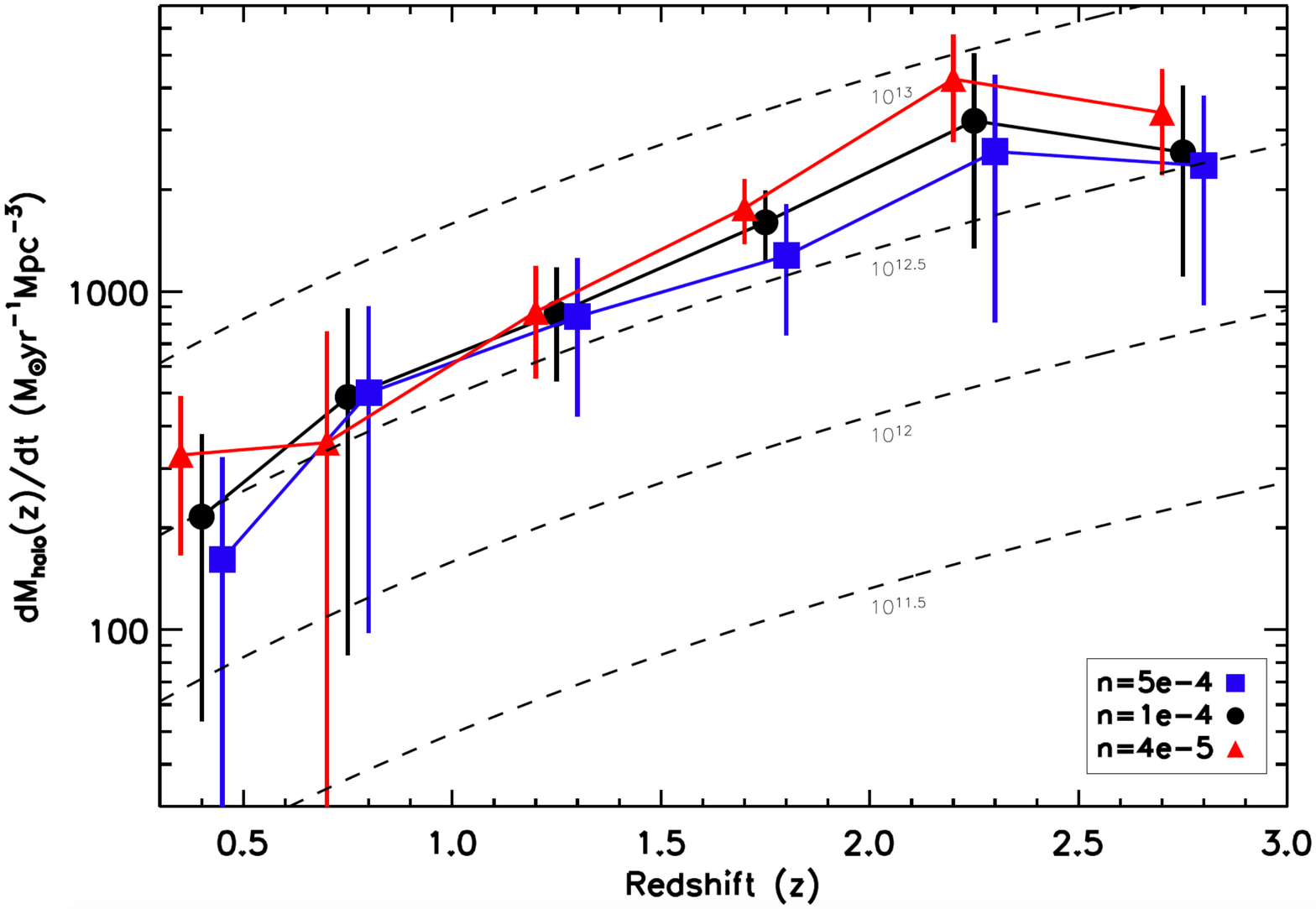}

          \caption{The derived dark matter accretion rate using the 
relations for stellar and halo masses which we derive in this paper.  These 
are plotted as a function of co-moving number density based on the star 
formation rate measures from Ownsworth
et al. (2014), such that lower number densities correspond to higher 
stellar masses.  
As can be seen rarer massive galaxies have a higher dark matter 
accretion rate
at higher redshifts, but that the rate of halo accretion appears to be 
similar within
this mass range.  Also plotted as the dashed lines are models for the
accretion rate of dark matter based on the theory of Correa et al. (2014)
with total masses for each model labeled.}

          \label{fig:BaryFrac}
        \end{figure}

\begin{equation}
\Psi(z) = \Psi_{0} \times {\rm exp} (-t/\tau)
\end{equation}

\noindent where $\Psi_{0}$ is the initial star formation rate; $t$ is the elapsed
time since the star formation rate began, and $\tau$ is the characteristic
declining rate of the exponential. This form is a good fit to the stellar populations of
galaxies at $z < 3$ (e.g. Mortlock et al. 2013; Ownsworth et al. 2014).   
 Integrating this between two times, $t_{1}$
and $t_{2}$, we obtain a measure of the amount of stellar mass created between
these two epochs:

\begin{equation}
\delta M_{*} = \tau \times {\rm \Psi_{0}} [{\rm exp}(-t_{1}/\tau) - {\rm exp}(-t_{2}/\tau)]
\end{equation}

\noindent if we take the star formation rate to begin at $t_{1} = 0$ then this
equation reduces to

\begin{equation}
\delta M_{*} = \tau \times \psi_{0} [1 - {\rm exp}(-t_{2}/\tau)].
\end{equation}

\noindent In what follows we consider as well the amount of stellar mass which
is returned to the ISM of these galaxies after taking into account the
results of stellar evolution after 1 Gyr (e.g., Conselice et al. 2013; Ownsworth
et al. 2014).  If we furthermore consider the form of the
ratio of stellar mass to total mass, i.e.,

\begin{equation}
f_{\rm *, halo}(z, M_{*}) = \frac{M_{*}(z)}{M_{\rm halo}(z)} 
\end{equation}

\noindent then a change in stellar mass $\delta M_{*}$ corresponds to
a change in the total mass given by

\begin{equation}
\delta M_{\rm halo} = \frac{1}{f_{\rm *,halo}} \times \delta M_{*}.
\end{equation}

\noindent  Where we have used the observation that $f_{*}$ is roughly
independent of redshift at $z < 0.5$.  Using this, we can then consider how the star 
formation rate relates to the
change in the stellar mass as given by eq. (26) over a large time period.
Between $t_{1}$ and $t_{2}$ the
addition of total mass to the system can be calculated by combining with
eq. (28).  We can then write,

\begin{equation}
\delta M_{\rm halo} = \frac{\tau \times \psi_{0} [1 - {\rm exp}(-t_{2}/\tau)]}{f_{*}} 
\end{equation}

\noindent however, we can also consider the instantaneous rate of accretion 
of material in terms of a star formation rate if we use the approximation 
that $\delta M_{*} \sim \psi(z) {\rm d}t$.  Using this we can then write 
the mass accretion rate as 

\begin{equation}
\frac{d M_{\rm halo}(z)}{dt} = \frac{1}{f_{*}} \psi^{'}(z) 
\end{equation}

\noindent where $\psi^{'}$ is the equivalent star formation rate after
accounting for the amount of mass returned to the ISM through stellar 
evolution processes.  Effectively, this is the stellar mass formation rate, 
rather than just the star formation rate.  Through the use of stellar 
evolution models the effective stellar mass formation rate is around
70\% of the star formation rate after a few Gyr.    

From Ownsworth et al.
(2014) we know how the star formation rate of galaxies 
with similar masses to 
those studied here evolves. Overall, the star formation rate is given by the form 
in eq. (24), with $\psi_{0} = 135$ \msol year$^{-1}$ and 
$\tau = 2.4 \times 10^{9}$ years.  Furthermore, the observed star 
formation rates are: $\psi = 55 $\msol year$^{-1}$ at $z = 3$; 
$\psi = 34$ \msol year$^{-1}$ at $z = 2$; and $\psi = 12$ \msol year$^{-1}$ at
$z = 1$.

Using the value of $f_{\rm *, halo}$ given in Section 5, and using eq. (30) 
we calculate the total mass accretion rate. We show 
these calculated values as a function of mass and redshift in Figure~20.  This 
figure reveals that the total mass accretion rate for typical galaxies 
varies with stellar mass and redshift. We find that the peak halo
accretion rate is around: $\dot{M}_{\rm halo} \sim 4000$ \msol year$^{-1}$ at $z \sim 2.5$.  This is much higher than the stellar mass or the
gas accretion rates at similar redshifts (e.g., Conselice et al. 2013; 
Mortlock et al. 2014; Ownsworth et al. 2014).

We find that within our assumptions galaxies grow their total integrated halo 
mass \Mtot\ at a self-similar rate as their stellar \Mst\  
masses.  The \Mst\ growth can be explained by the process of 
mergers (both major and minor) triggering star-formation events which 
increases \Mst. The gas for this 
must come from both minor/major 
mergers, as well as gas accretion (e.g., Conselice et al. 2013; Mundy et al. 2017).  
The fact that the dark matter accretion rate follows that of the star 
formation rate, revealing a 
nearly constant ratio at $z > 1$, implies that the dark matter is being accreted 
with the 
baryons in these galaxies to build up the mass of the system at this epoch.

We can also compare these values to theory which predicts, even based on
analytical calculations, what the accretion rate of dark matter is onto
halos (e.g., van den Bosch et al. 2002; Wechsler et al. 2002; McBride
et al. 2009; Correa et al. 2014).   For example, Wechsler et al. (2002) predict an
accretion history which scales as an exponential: M$\sim$e$^{- \alpha z}$.

It is proposed based on these various simulations and calculations using
the Extended Press-Schechter (EPS) formalism that the total mass accretion
history can be represented by a combination of an exponential and a power-law
in the form (e.g., McBride et al. 2009; Correa et al. 2014):

\begin{equation}
M(z) = M_{0} \times (1+z)^{\alpha} \times {\rm e}^{\beta \times z}
\end{equation}

\noindent  Taking the derivative of this, we find that the rate of change of
the halo mass is given by (e.g., Correa et al. 2014)

\begin{eqnarray}\nonumber
\frac{dM}{dt} &=& 71.6 \rm{M}_{\sun}\rm{yr}^{-1} M_{12}\times\\\nonumber
&& [-\alpha-\beta(1+z)][\Omega_{m}(1+z)^{3}+\Omega_{\Lambda}]^{1/2}.
\end{eqnarray}

\noindent where ${\rm M_{12}}$ is the halo mass of the galaxy in units of
$10^{12}$ \msol, and $\Omega_{\rm m}$ and $\Omega_{\lambda}$ are the 
cosmology.  

When we fit our data and observations to this we find that the values of
$\alpha$ and $\beta$ are as given in Table~10.  This is in agreement with our
observations, as can be seen in the comparison in Figure~20.   This is similar
to the results of the mass accretion rates derived from other, but similar, 
assumptions (e.g., e.g., Behroozi et al. 2013; Rodriguez-Puebla et al. 2017).


        \begin{table}
          \begin{tabular}{c c c  c c}
            \hline
            Density (Mpc$^{-3}$) & log M$_{*}$ & log M$_{\rm halo}$ & $\alpha$ & $\beta$ \\
            \hline
5 $\times 10^{-4}$ & 10.8 & 12.4 & -0.44$\pm$1.25 & -0.88$\pm$0.41 \\
1 $\times 10^{-4}$ & 11.2 & 13.1 & -0.12$\pm$0.37 & -0.20$\pm$0.11 \\
4 $\times 10^{-5}$ & 11.4 & 13.3 & 0.07$\pm$0.33 & 0.21$\pm$0.11 \\
            \hline
          \end{tabular}
          \caption{The values of $\alpha$ and $\beta$ for galaxies selected at different number densities (see Ownsworth et al. 2014).   These values give the form for the growth of halos at the given mass limit and are consistent with the theoretical work of Correa et al. (2014)}
        \label{tab:alphabeta}
        \end{table}

\subsection{Dark Matter Mass as a Regulator of Galaxy Formation }

One of the goals in this paper is to relate the matter components of galaxies beyond the 
local universe, and to develop a method for doing this which can be applied to single galaxies, 
as opposed to stacked systems as used in clustering and lensing (e.g., Coupon et al. 2015).     
In general we find that the stellar mass is, within some scatter, a good tracer of the 
dynamical mass and halo masses of field galaxies at $z < 1.2$ within our mass range.   
This may not be the case for red passive galaxies, or for those in very dense areas such as
clusters, which are not present in our sample 
in great abundance. 

This suggests that the different masses in galaxies, and how they build up over time, are highly regulated 
by the overall halo mass of the galaxy which drives the formation of the amount of stellar mass within
these galaxies. That is, fundamentally
the formation of galaxies is hierarchical with dark matter accreted from the IGM the main way mass is built up.  When this dark matter is accreted baryonic mass is as well, which leads to additional galaxy formation processes such as star formation and likely AGN activity.  This also suggests that galaxies are assembling 
hierarchically, independent of the method in which mass is brought into the systems, whether it be 
through mergers of various types or through e.g., gas accretion processes.  

This is similar to previous findings from other papers, including Kravtsov et al. (2014) where the masses of
brightest cluster galaxies (BCGs) have a stellar mass which scales with total mass (M$_{500}$) such that 
M$_{\rm *, BCG} \sim$ M$^{0.4}_{500}$ with a scatter in M$_{500}$ of 0.2 dex for centrals.  Satellite galaxies
within this sample have s scaling which goes as M$_{\rm *, sat} \sim$ M$^{0.8}_{500}$ with a smaller
scatter of 0.1 dex. This is similar to our results when we examine the scaling between the stellar mass
and halo mass for our systems (\S 5.1.2).

\subsection{Galaxy Formation Efficiency}

A standard way to characterize the efficiency of galaxy formation is to 
compare the stellar masses of galaxies to their halo masses, and to find where
this ratio peaks as a function of stellar mass and redshift.  Furthermore it is
interesting to determine if this peak changes as a function of redshift, as well 
as if different methods give different answers to this observational question.

It is well known that the peak of the galaxy formation efficiency, i.e., when 
the  stellar to dark matter masses peaks, is at roughly log $M_{*} = 10.5$ 
or log M$_{\rm halo} \sim 12$ (e.g.,
Behroozi et al. 2013) through the use of abundance matching.  Our results show a few things that are worth discussing in detail and following up, as we do not find such clear cut evidence for a turn over, at least to within the mass limits of our sample.

First, we find that the ratio of stellar mass to halo mass declines as we observe galaxies at 
lower stellar masses at $z < 1.5$.  This is as expected,
and shows that the galaxy formation process efficiency is lower for lower mass galaxies, as seen 
in the comparison points from abundance matching and lensing as shown in Figure~16.  However, 
we do not find a `peak' in the galaxy formation process, as we find a continual increase in the 
ratio of stellar to halo mass for individual galaxies as we go to higher masses.  This suggests 
that the dark matter masses we are measuring are only for single galaxies, and not large 
halo masses as what might be measured in groups or clusters of galaxies. Our samples are generally 
not taken from dense environments, and this may be one reason we do not see an enlarged halo mass 
for the highest mass galaxies.

There is also the issue that the peak in the galaxy formation may evolve
with redshift, as is found in halo clustering derived masses (e.g., 
Coupon et al. 2012). However,
this is likely due to the fact that the halo masses derived with clustering
measures  are
observing more the massive parent halo masses due to halo downsizing.  We do not see any evidence for
an evolution in the ratio of stellar to halo mass, nor do we see evidence that the galaxy formation efficiency
peak is changing at higher redshifts.    

Based on current ideas, we would expect the stellar mass to halo mass ratios for the most
 massive galaxies to decline from a peak at the highest masses.  We would see this as the most massive
galaxies having the largest relative dark matter content.  This is usually interpreted as being due to AGN
feedback which is most effective at the highest stellar and halo masses which prevents gas
from cooling to form stars.  The fact that we do not see this, at least when we compare this
ratio with stellar mass, suggests that halo masses for these very massive galaxies we are measuring are for the halo associated with just the galaxy itself, and not the overall larger halo that might be present.  This may alternatively imply that the effects of AGN feedback are not important for determining the total amount of baryonic mass formed, or that the effects are most pronounced in the outer parts of halos. 

In fact, the greatest effect we see is a form of halo downsizing, whereby the earliest galaxies to form in the `oldest' halos have the largest stellar mass to dark matter mass. This implies that the time these halos forms is more critical for their overall `efficiency' of galaxy formation than any other process. We can see this with a higher stellar mass to halo mass ratio at a given mass for systems which are ellipticals compared with for example disks.  This higher stellar to halo mass ratio is also seen in the models, suggesting that the time of formation is indeed the third parameter in the correlation between the stellar mass and halo masses of galaxies which correlates with the dynamical mass (\S 4.3).

  \section{Conclusions \& Summary}
    \label{sec:Conclusion}
    
In this paper we examine the different forms of galaxy mass (stellar, dynamical, halo) at high redshifts from $z = 0.4 - 3$.  We take as our starting point a well defined sample of 544 galaxies with well measured star formation rates, kinematics and sizes and derive relations between the derived gas masses, dynamical masses and halo masses.  We find the following major results.

\noindent 1. Galaxy halo and dynamical masses can be successfully retrieved used the S$_{0.5}$ index which combines the use of a galaxy's velocity dispersion and its rotational velocity.   This is superior to using just a velocity dispersion or a rotational velocity and allows the measurement of halo masses on individual systems as opposed to ensemble averages.  We furthermore argue that the third parameter in this fit is the time of halo formation with systems forming earlier having a higher stellar to halo mass ratio.

\noindent 1. There is a strong relation between the stellar mass and dynamical masses for galaxies that occupy their own massive halos (i.e.,
field galaxies).  This suggests that galaxy formation is highly regulated and the different masses are correlated and assemble together in galaxies.

\noindent 2. We show that within the visible radius, dynamical masses of the galaxies in our sample are dominated by stellar mass up to $z = 1.2$.  

\noindent 3. We derive the halo masses of galaxies by using semi-analytical models which show how to relate the velocity of rotation, size and halo mass together.  This method shows that the most massive galaxies in stars have the highest ratios of stellar to halo mass, while the least massive galaxies have the lowest ratios.

\noindent 4. We develop a series of empirically based relations between the stellar mass of galaxies and their  dark masses as a function of redshift.   The scatter in these relations are just slightly higher than the expected error budget in stellar masses, with a dispersion of around $\sim 0.3$ dex. 

\noindent 5.  We use the fact that the stellar to halo mass does not appear to 
change significantly with redshift at $z > 1$ to argue that the dark matter accretion rate is 
$\dot{M}_{\rm halo} \sim 4000$ \msol year$^{-1}$ at $z \sim 2.5$ down to 
a accretion rate of a few hundred \msol year$^{-1}$ by $z \sim 0.5$.

In the future, these types of observations can be extended with longer exposures obtaining deeper kinematic data in the outer parts of galaxies which will hopefully give us a better and more accurate idea of the halo masses for these systems.  Doing this is however very difficult with current technology, and may require the use of spectrographs on 20-30m telescopes in the next decade.  In the mean-time clustering and lensing analyses of masses will make headway in terms of average galaxy properties and we can measure mass evolution for bulk galaxies divided into finer subsets.  

  \section*{ACKNOWLEDGMENTS}
 
We thank the referee for their comments which significantly improved the presentation of this paper.  We thank Alice Mortlock, Jamie Ownsworth and Seb Foucaud for their comments and help with various aspects of this paper, and Andrew Benson for help with the use of the Galacticus models.  We thank Sedona Price and the rest of the MOSDEF team for making their data available, and to Alessandra Beifiori for making her SDSS-III galaxy internal measurements of sizes and velocities available.  
We acknowledge support from the Leverhulme Trust in the form of a Leverhulme Prize to CJC and funding from the STFC as well as financial support from the University of Nottingham.

\appendix

\section{The Meaning and Usefulness of the S$_{0.5}$ Parameter}

As mentioned in the body of the paper, we use the S$_{0.5}$ parameter (\S 3.2) 
throughout to derive the dynamical masses of
galaxies, and furthermore we use this measurements as a proxy and as a way to obtain the halo masses of a galaxy.  We investigate
in this appendix the nature of the S$_{0.5}$ parameter and how to interpret the resulting 
dynamical mass calculated from it.    

First, we determine how the distribution of S$_{0.5}$ values in the models match the data.  To do this, we compare at both $z < 1$ and $z > 1$ the distribution of half-light radii and S$_{0.5}$ values.  As shown in Figure A.1 there is a good overlap in these values, i.e., that the simulation we use to calculate the dynamical masses has a good overlap with the actual data at both high and low redshift.   This is also the case when comparing with the observed and
predicted relation between S$_{0.5}$ and stellar mass (Figure~10; \S 5.1.1) in a Tully-Fisher like
relation. 

The one exception to this is that we observationally do not see high velocity systems which have very small radii in the same abundance as the simulations.    We do however find that there are galaxies with larger radii, namely the early-types, which have large S$_{0.5}$ parameters but which are not abundant in the model itself.   We therefore exclude these simulated galaxies in our further analyses which have sizes $> 10$ kpc.   We also find systems which have a very small radii and a large value of combined velocities through the S$_{0.5}$ parameter which we also exclude from our analyses.

          \begin{figure}
          \centering
\vspace{-1cm}

          \includegraphics[angle=-90, width=17cm]{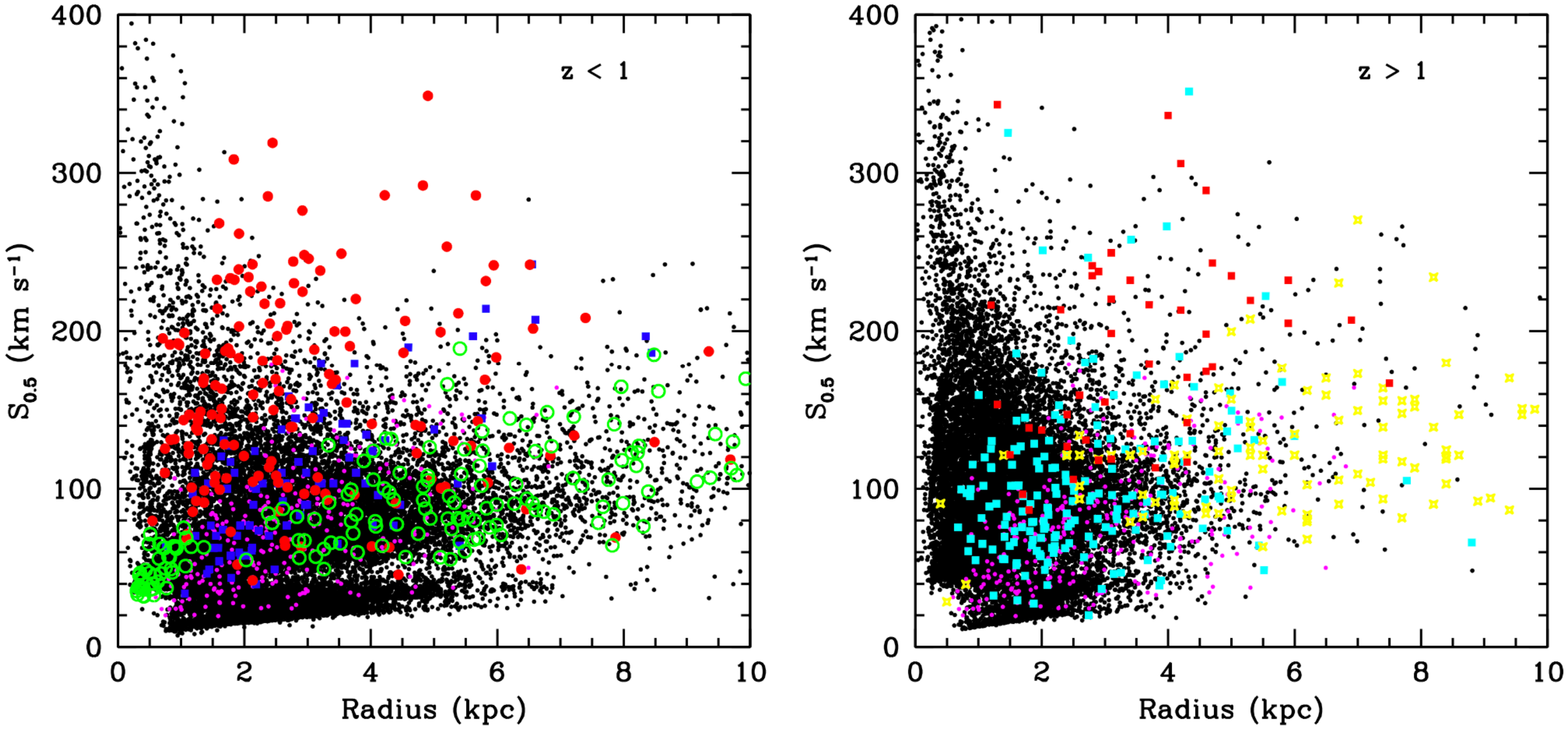}
\vspace{-2cm}

\caption{The relationship between the S$_{0.5}$ parameter and the sizes of galaxies, both of which go into calculating the dynamical masses.  The small black points are from the galacticus simulation, while the remaining points are the same as in Figure~11. }
        \end{figure}

The nature of dynamical masses from S$_{0.5}$ is such that it should 
represent a good indicator for the total mass within the galaxy itself, or that it correlates in some way with the total amount of mass within the galaxy.  Thus, within the
brighter portion of a galaxy, within R$_{e}$, we would expect that the total mass of our galaxies are dominated by
the stellar mass,  as dark matter would have very little expected contribution to the inner portions of galaxies.  

To demonstrate how well this assumption is we plot at three different redshifts the relationship
between the dynamical mass and the stellar mass in Figure~A2.  Here the dynamical mass is measured at exactly the half-light radius, and thus we do not implement the $\beta$ term from eq. 3 here.

          \begin{figure*}
          \centering
\vspace{-3cm}

          \includegraphics[angle=-90, width=19cm]{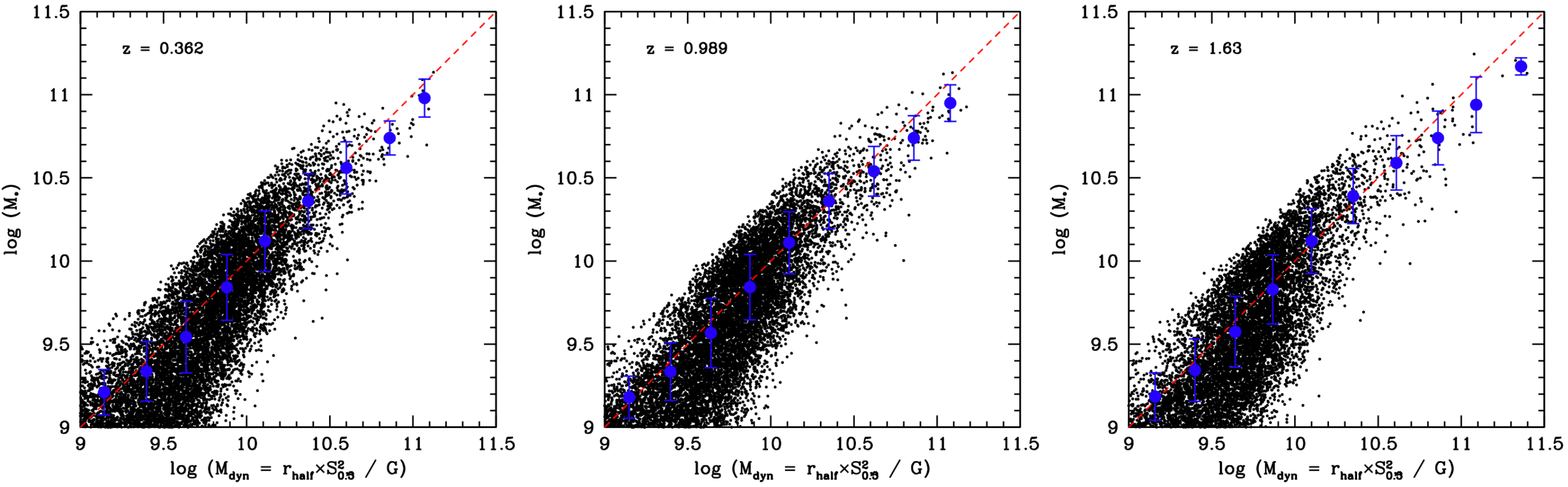}
 \vspace{-3.5cm}

         \caption{The relationship between the dynamical mass and the stellar mass, with the
dynamical mass defined using the S$_{0.5}$ parameter as in eq. 2.  The dashed red line shows
the 1:1 relation between these two quantities and the blue points with error bars show the
average values at different masses and the scatter within that dynamical mass bin. }
          \label{fig:MvirMstMbary}
        \end{figure*}

As can be seen, there is a good relationship between the stellar mass and the dynamical
mass within the Galacticus models.  This relationship is nearly 1:1 up to about stellar masses
of M$_{*} = 10^{11}$ \solm when the relationship becomes softer, particularly at higher dynamical
masses at a lower redshifts. However, this difference is less striking at higher redshifts, $z = 1.63$, where
the relationship between these two quantities is slightly improved.    This figure however does justify the use of our dynamical mass as a measure of 'mass' in a galaxy.  Given that it is kinematically based, and measured within the visible portions of galaxies, we find that it statically matches the stellar mass at the location where the stellar mass dominates the total mass of a galaxy.

\section{Other Fitting Forms}

\subsection{Modeled Stellar Mass to Halo Mass Relation}

There are a myriad of ways that the relation between the stellar mass or dynamical mass
can be fit as a function of the halo mass.  In this section we discuss some of these fits
and give their forms. We also consider how other masses, the circular mass and the sigma mass,
are related to the halo mass. First, we consider however alternative methods for measuring the
halo mass based on the dynamical mass.

The first relationship we consider is whereby the relation between stellar to halo mass is
fit by two power-laws with a break at M$_{*} \sim 10^{9.9}$ \msol which is the representative
location where the slope of the relationship changes.  We fit 
this relation between stellar and halo mass with a linear relationship between the log of these 
quantities using the form:

\begin{equation}
{\rm log M_{\rm halo}} = \alpha \times {\rm log M_{*}} + \beta
\end{equation}

\noindent where the quantities for $\alpha$ and $\beta$ are listed in Table~B1 at the separation point
log M$_{*} = 9.5$.   As we find in the main paper we find that there is no evolution in the relationship between stellar and halo masses, at least in these models.

        \begin{table}
          \begin{tabular}{c c c c c}
           
                     & log M$_{*} < 9.9$ &  &  log M$_{*} > 9.9$ & \\
            \hline
            Redshift & $\alpha$  & $\beta$ & $\alpha$ & $\beta$  \\
            \hline
              0.4  & 0.44$\pm$0.04 & 6.8$\pm$0.3 & 1.68$\pm$0.12 & -5.8$\pm$1.3 \\
              0.7  & 0.38$\pm$0.05 & 7.5$\pm$0.4 & 1.65$\pm$0.05 & -5.5$\pm$0.6 \\
              1.0  & 0.40$\pm$0.05 & 7.2$\pm$0.4 & 1.65$\pm$0.04 & -5.4$\pm$0.5 \\
              1.3  & 0.44$\pm$0.05 & 6.9$\pm$0.5 & 1.58$\pm$0.02 & -4.7$\pm$0.2 \\
              1.6  & 0.44$\pm$0.05 & 6.9$\pm$0.4 & 1.55$\pm$0.02 & -4.2$\pm$0.1 \\
              1.9  & 0.42$\pm$0.04 & 7.1$\pm$0.4 & 1.53$\pm$0.04 & -4.0$\pm$0.4 \\
            \hline
          \end{tabular}
        \caption{The fitted values of $\alpha$ and $\beta$ for the relation between stellar and halo mass as found through the Galacticus simulation. These values are used in Equation (B1).  }
        \label{tab:stellarhalo}
        \end{table}

We also examine results from the Millennium simulation to see if there is any differences between different models for the relationship between stellar and halo mass.   To carry out this comparison  we select galaxies from the De Lucia \& Blaizot (2007) catalogue at $z\sim 1$ giving us 684,357 galaxies selecting both stellar and halo masses.  We use the friend-of-friend halo masses from this simulation as a measure of our halo mass.   These friends-of-friends masses are not based on a certain mass threshold, but on a linking length parameter, and there is a variation in the mass overdensity depending upon the concentration of the various mass profiles (e.g., More et al. 2011).   The corresponding relationship for the Millennium simulation over our redshifts of interest is given by:

         $${\rm log} \mbox{M}_{\rm{halo}} = (0.584 \pm 0.029) \, \log \left(\mbox{M}_{\rm{*}}\right) + (5.556 \pm 0.271)$$

 \begin{equation}
         {\mbox{M}_{\rm{*}} < 10^{9.9}} 
\end{equation}
$$          {\mbox{log M}_{\rm{halo}} = (1.442 \pm 0.060) \, \log \left(\mbox{M}_{\rm{*}}\right) - (3.107 \pm 0.616)}. $$
 $$         {\mbox{M}_{\rm{*}} \ge 10^{9.9}}. $$
          \label{eqn:fit_all}

\noindent Where we find a very similar relationship that we find in the Galacticus simulations, which we discuss in the main body of the paper.  

\subsection{Alternative Dynamical Mass-Halo Mass Relations}

\subsubsection{Moster et al. (2010) Functional Form}

One of the functional forms in which we fit the dynamical to halo mass is through that established
by Moster et al. (2010) for the relationship between the stellar mass and the halo mass through
abundance matching.  The form of this fitting for our relation between the dynamical and halo
mass is:

\begin{equation}
{\rm log\,} M_{\rm halo} = 2 \times ({\rm log\, M_{halo}})_{0}  \left[ \left( \frac{{\rm log} M_{*}}{M_{1}} \right)^{-\beta} + \left( \frac{{\rm log M_{*}}}{M_{1}} \right)^{\Gamma} \right].
\end{equation}

\noindent As discussed in \S 4.3 the best fitting relations for the Moster et al. (2010) relations is shown
as the green line in Figure~5.  As also discussed, these fits are not superior to the simplified form
of a single power-law relationship between the value of log M$_{\rm dyn}$ and log M$_{\rm halo}$.  We also use this form of fit when fitting the sigma mass to the halo mass in \S 4.3.3 and Figure~7.  It's likely that this more complex form is more accurate for fitting functional forms, something which can be explored in more detail once models predicting the quantities we study are better calibrated.

\allauthors

\listofchanges

\end{document}